\documentclass[preprint]{emulateapj}
\usepackage{apjfonts}

\newcommand{\galex}{{\it GALEX}}

\newcommand{\mic}{{$\mu$m}}
\newcommand{\lir}{{L_{\rm IR}}}

\slugcomment{2009 May 15}
\journalinfo{Accepted to ApJ}

\shorttitle{MID-IR LUMINOSITIES AND UV/OPTICAL SFRS}
\shortauthors{SALIM ET AL.}

\begin{document}

\title{Mid-IR Luminosities and UV/Optical Star Formation Rates at $z<1.4$} 
\author{Samir Salim\altaffilmark{1},
Mark Dickinson\altaffilmark{1},
R.\ Michael Rich\altaffilmark{2},
St\'ephane Charlot\altaffilmark{3},
Janice C.\ Lee\altaffilmark{4},
David Schiminovich\altaffilmark{5},
Pablo G.\ P\'erez-Gonz\'alez\altaffilmark{6,7},
Matthew L.\ N.\ Ashby\altaffilmark{8},
Casey Papovich\altaffilmark{7,9},
S.\ M.\ Faber\altaffilmark{10},
Rob J.\ Ivison\altaffilmark{11},
David T.\ Frayer\altaffilmark{12},
Josiah M.\ Walton\altaffilmark{13},
Benjamin J.\ Weiner\altaffilmark{7},
Ranga-Ram Chary\altaffilmark{14},
Kevin Bundy\altaffilmark{15},
Kai Noeske\altaffilmark{8},
Anton M.\ Koekemoer\altaffilmark{16}}

\altaffiltext{1}{National Optical Astronomy Observatory, 
  950 North Cherry Ave., Tucson, AZ 85719, samir@noao.edu}
\altaffiltext{2}{Department of Physics and Astronomy, University of
  California, Los Angeles, CA 90095}
\altaffiltext{3}{Institut d'Astrophysique de Paris, CNRS, 98 bis
  boulevard Arago, F-75014 Paris, France}
\altaffiltext{4}{Carnegie Observatories, 813 Santa Barbara
  Street, Pasadena, CA  91101}
\altaffiltext{5}{Department of Astronomy, Columbia University,
  New York, NY 10027}
\altaffiltext{6}{Departamento de Astrof\'{\i}sica, Facultad de CC. 
  F\'{\i}sicas, Universidad Complutense de Madrid, E-28040 Madrid, Spain}
\altaffiltext{7}{Steward Observatory, University of Arizona, Tucson, AZ 85721}
\altaffiltext{8}{Harvard-Smithsonian Center for Astrophysics,
  60 Garden Street, Cambridge, MA 02138, USA}
\altaffiltext{9}{Department of Physics, Texas A\&M University, 
College Station, TX 77843}
\altaffiltext{10}{University of California Observatories/Lick Observatory, 
University of California, Santa Cruz, CA 95064}
\altaffiltext{11}{UK Astronomy Technology Centre, Royal Observatory, 
  Blackford Hill, Edinburgh EH9 3HJ, United Kingdom}
\altaffiltext{12}{Infrared Processing and Analysis Center, California
Institute of Technology 100-22, Pasadena, CA 91125}
\altaffiltext{13}{University of Arkansas, Fayetteville, AR 72701}
\altaffiltext{14}{Spitzer Science Center, California Institute of
Technology 220-6, Pasadena, CA 91125}
\altaffiltext{15}{Dept.~of Astronomy and Astrophysics, University of Toronto, 
50 St.~George Street, Toronto, ON M5S 3H4, Canada}
\altaffiltext{16}{Space Telescope Science Institute, 3700 San Martin
Drive, Baltimore, MD 21218}

\begin{abstract}
Ultraviolet non-ionizing continuum and mid-IR emission constitute the
basis of two widely used star formation indicators at intermediate and
high redshifts. We study 2430 galaxies with $z<1.4$ in the Extended
Groth Strip with deep MIPS 24 \mic\ observations from FIDEL,
spectroscopy from DEEP2, and UV, optical, and near-IR photometry from
AEGIS. The data are coupled with dust-reddened stellar population
models and Bayesian SED fitting to estimate dust-corrected SFRs. In
order to probe the dust heating from stellar populations of various
ages, the derived SFRs were averaged over various timescales--from 100
Myr for ``current'' SFR (corresponding to young stars) to 1--3 Gyr for
long-timescale SFRs (corresponding to the light-weighted age of the
dominant stellar populations). These SED-based UV/optical SFRs are
compared to total infrared luminosities extrapolated from 24 \mic\
observations, corresponding to 10--18 \mic\ rest frame. The total IR
luminosities are in the range of normal star forming galaxies and
LIRGs ($10^{10}$--$10^{12} L_{\odot}$). We show that the IR luminosity
can be estimated from the UV and optical photometry to within a factor
of two, implying that most $z<1.4$ galaxies are not optically
thick. We find that for the blue, actively star forming galaxies the
correlation between the IR luminosity and the UV/optical SFR shows a
decrease in scatter when going from shorter to longer SFR-averaging
timescales. We interpret this as the greater role of intermediate age
stellar populations in heating the dust than what is typically
assumed. Equivalently, we observe that the IR luminosity is better
correlated with dust-corrected optical luminosity than with
dust-corrected UV light. We find that this holds over the entire
redshift range. Many so-called green valley galaxies are simply
dust-obscured actively star-forming galaxies. However, there exist 24
\mic-detected galaxies, some with $\lir>10^{11}L_{\odot}$, yet with
little current star formation. For them a reasonable amount of dust
absorption of stellar light (but presumably higher than in nearby
early-type galaxies) is sufficient to produce the observed levels of
IR, which includes a large contribution from intermediate and old
stellar populations. In our sample, which contains very few ULIRGs,
optical and X-ray AGNs do not contribute on average more than $\sim
50\%$ to the mid-IR luminosity, and we see no evidence for a large
population of ``IR excess'' galaxies.

\end{abstract}

\keywords{galaxies: evolution---galaxies: fundamental parameters---
infrared: galaxies---ultraviolet: galaxies---surveys---galaxies:
active}

\section{Introduction} \label{sec:intro}

The total infrared (IR) luminosity, either alone or in combination
with the ultraviolet (UV) luminosity \citep{heckman98}, is
increasingly being considered a reliable star formation (SF) indicator
for normal, dusty star-forming galaxies \citep{kewley02}. This is
especially the case since the more traditional SF\footnote{SF will be
used to designate ``star formation'' or ``star forming'', depending on
the context.}  indicators, such as the UV continuum and nebular line
flux, require somewhat substantial corrections for dust extinction
\citep{kennicutt}. The {\it mid}-infrared luminosity has recently been
suggested as a tracer of star formation
\citep{roussel,forster,wu,alonso,calzetti07,rieke}, potentially
serving as an alternative to the far IR, which is more difficult to
obtain. The mid IR has received particular attention in intermediate
and high redshift studies, largely driven by the sensitivity of {\it
Spitzer} MIPS observations, which with its 24 \mic\ detector readily
observes normal star forming galaxies ($\lir \sim 10^{10} L_{\odot}$)
out to $z\sim 1$ (e.g., \citealt{lefloch}) and luminous and
ultra-luminous IR galaxies (LIRGs, ULIRGs) out to $z\sim 2$ (e.g.,
\citealt{papovich,reddy}).

The validity of using the IR as a SF indicator at intermediate
redshifts depends critically on the assumption that the IR flux is
tightly correlated with young stellar populations for {\it typical}
field galaxies in deep surveys. While one expects dust-reprocessed
emission from both young and old stars to contribute to the IR, the
question of a dominant source is less straightforward. The source of
the {\it far-IR} emission in nearby star forming galaxies has been a
subject of debate predating the launch of {\it Spitzer Space
Telescope}. That the majority of IR heating is due to young
populations, i.e., hot stars located in compact star forming regions,
has been initially suggested by the similarity between H$\alpha$ and
far-IR structures within nearby galaxies (e.g.,
\citealt{dev97}). Studies utilizing better resolution from {\it
Spitzer} to some degree confirmed these earlier findings and extended
them down to 70 and 24 \mic\ \citep{hinz,pablo_m81}. On the other
hand, the claims for a more significant role of older stellar
populations in the far IR, which heat the dust through a diffuse
interstellar radiation field, were initially based on the modeling of
\citet{walterbos}, who successfully predicted {\it IRAS} 60 and 100
\mic\ fluxes using dust models and assuming that $B$-band light (from
intermediate age stars; $\sim 1$ Gyr) traces the general interstellar
radiation field. While it is now generally accepted (e.g.,
\citealt{dacunha}) that the interstellar radiation field can be a
significant heating source for the far IR, \citet{boselli} suggested
that this may be true for the {\it mid IR} as well. They found that
6.75 and 15 \mic\ emission measured by {\it ISO} correlates better
with far-IR luminosity than with either H$\alpha$ or UV dust-corrected
luminosity. More recently, the case for the interstellar radiation
field producing the 8 \mic\ PAH emission has been made by
\citet{bendo} who find a good correlation with 160 \mic\ emission. On
the other hand, \citet{diaz} find that 8 \mic\ emission from HII
regions in local LIRGs follows Pa$\alpha$ emission from young stars
when metallicity and age are fixed. However, unlike the emission at 8
\mic, the general consensus for mid-IR continuum at 24 \mic\ is that
it is dominated by emission from star-forming regions
\citep{calzetti07,rieke}.

The goal of this study is to explore the use of mid-IR luminosity
(specifically in 10--18 \mic rest-frame range) as a SF indicator. This
wavelength range falls inbetween the 8 \mic\ IRAC and 24 \mic\ MIPS
bands, where there are no direct constrains from {\it Spitzer} studies
of nearby galaxies. Also, our sample of 24 \mic-detected galaxies at
$0.2<z<1.4$ is generally more luminous than the samples studied
locally (such as SINGS). We base our approach on the comparison of the
level of correlation between total IR luminosities (extrapolated from
MIPS 24 \mic\ observations) and UV/optical dust-corrected star
formation rates (SFRs). These SFRs come from UV/optical SED fitting,
which allows us to construct SFRs averaged over various timescales,
from 0.1 to several Gyr. SFRs averaged over various timescales
correspond to dust-corrected luminosities coming from stellar
populations ranging in age. We perform the comparison for various
subsamples, specifically for blue actively star-forming galaxies and
red quiescent ones. Finding the age of the stellar population that
best correlates with IR luminosity could indicate the stellar
population responsible for dust heating at 10--18 \mic. In \S
\ref{sec:data} we present the multiwavelength data sets used in this
study. In \S \ref{sec:sed} we derive SFRs from UV/optical SED fitting,
and in \S \ref{sec:lir} we derive IR luminosities from 24 \mic\
observations. The results of the comparison of UV/optical SFRs and IR
luminosities of blue star forming galaxies are presented in \S
\ref{sec:lir_sfr}, while red (dusty or quiescent) galaxies and AGN
candidates are analyzed in \S \ref{sec:lir_red}.  In this paper we use
a $\Omega_m=0.3$, $\Omega_\Lambda=0.7$, $H_0= 70\, {\rm km\, s^{-1}\,
Mpc^{-1}}$ cosmology.

\section{Data} \label{sec:data}

In this study we use various data sets matched to the DEEP2 redshift
survey. Redshifts and UV, optical, and $K$-band photometry are part of
the All-Wavelength Extended Groth Strip International Survey (AEGIS,
\citealt{aegis}). AEGIS combines observations from a number of
ground-based and space observatories.\footnote{Please refer to
\url{http://aegis.ucolick.org} for more information on AEGIS,
including the footprint of various data sets.} The DEEP2 sample is
$R$-band selected, and we maintain this selection by keeping all
objects even if they are not matched with certain bands. In most of
the paper we study the subset of this optical sample that is detected
at 24 \mic. Therefore, one has a combination of $R$-band and 24 \mic\
selections. 24 \mic\ data come primarily from the Far Infrared Deep
Legacy (FIDEL) survey. Main properties of the data sets are given in
Table \ref{tab:data}.

\subsection{DEEP2 redshifts}

The core data set to which we match all other data is the DEEP2
redshift survey of the Extended Groth Strip (EGS), one of the four
fields of the full DEEP2 survey \citep{deep2,faber}. DEEP2 EGS spectra
form the basis of the AEGIS survey. They were obtained with the DEIMOS
spectrograph on Keck II and cover a wavelength range 6400-9100 \AA\
with 1.4 \AA\ resolution. We use the 2007 version of the redshift
catalog containing 16087 redshifts, of which 10743 are considered
secure (quality flag 3 or 4), representing a 13\% increase over the
catalog described in \citet{aegis}. Galaxies were optically selected
to be brighter than $R=24.1$\footnote{Magnitudes are given in AB
system throughout.}, with a known selection function, resulting in a
redshift distribution with a mean redshift of 0.7 and extending up to
$z\sim1.4$ \citep{faber}.

\subsection{\galex\ UV photometry}

\begin{figure}
\epsscale{1.0}
\plotone{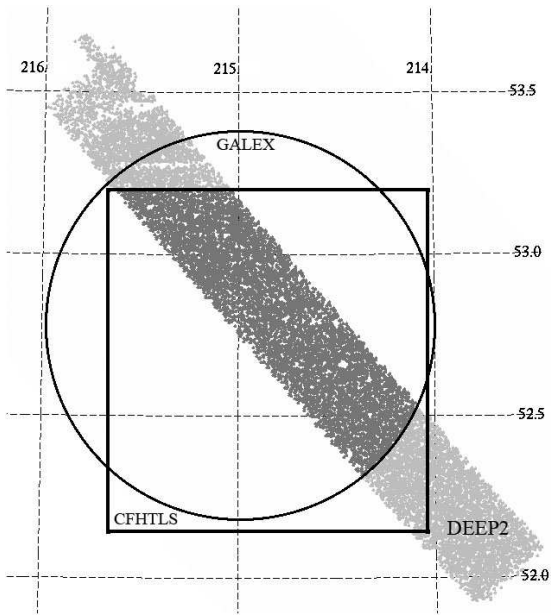}
\caption{Areal map of the sample. Our optical sample consists of DEEP2
galaxies (gray region) having secure spectroscopic redshifts and
lying in the intersection of {\it GALEX} (circle) and CFHTLS
(rectangle) areas (dark gray, 0.31 sq.\ deg). Other surveys used in
the SED fitting ($u$-band MMT and $K$-band Palomar) cover the dark
gray area almost entirely and their footprints are not shown
here. Note that a galaxy is kept in the optical sample even if it is
not detected in all bands, as long as it lies in the sample area (dark
gray region). Thus, the optical sample is only $R$-band selected. We
derive SED fitting parameters for the entire optical sample, but then
study a subset of objects that are detected at 24 \mic. The 24 \mic\
data (footprint not shown) cover the dark gray region fully.}
\label{fig:area}
\end{figure}

\galex\ \citep{galex,morrissey} imaged the central portion of the EGS
with a single $1.2\degr$ diameter pointing (Figure
\ref{fig:area}). The exposure time was 237 ks in the near-UV (NUV) and
120 ks in the far-UV (FUV) band, which makes it the deepest \galex\
single pointing to date. Data are taken from public release GR3. While
\galex\ observes in FUV and NUV simultaneously, 1/2 of the FUV
exposure time was lost due to anomaly with the FUV detector
\citep{morrissey}. The \galex\ pipeline produces catalogs using
SExtractor \citep{sext} aperture photometry. While adequate for more
shallow, resolved images, such photometry suffers from severe blending
and source confusion in the deep EGS images (\galex\ resolution is
$4-5''$, while astrometry is good to $0\farcs5$,
\citealt{morrissey}). To remedy this problem we perform PSF source
extraction, which is less sensitive to blending \citep{zamojski}. We
use the custom-built PSF extraction software {\it EM Photometry},
developed by D.\ Aymeric, A.\ Llebaria and S.\ Arnouts, which uses the
expectation-maximisation (EM) algorithm of \citet{em}. {\it EM
Photometry} extracts \galex\ fluxes based on optical prior
coordinates. While successfully dealing with blending, the resulting
fluxes for a given set of objects will to some extent depend on the
depth (i.e., the number of objects) in the prior catalog, with
photometric bias being especially pronounced for intrinsically fainter
objects. In particular, having too many faint optical priors (fainter
than the equivalent \galex\ limit in the absence of blending) will
result in the splitting of the UV flux of one object among multiple
sources, most of which are not actually detectable in \galex\
images. We minimize this bias by using the list of optical priors
based on $u$-band (band closest to NUV) photometry from the CFHT
Legacy Survey (\S \ref{ssec:cfhtls}) and choosing a $u$ limit of 25.5,
at which the majority of optical objects still have real counterparts
in the NUV image. After extracting the NUV fluxes using the $u$-band
prior catalog, we find that genuine detections mostly have NUV$<26.5$,
which we adopt as a cut for the NUV catalog, and which roughly
corresponds to 3$\sigma$ limit. As a cross-check we also perform
source detection and PSF extraction using {\it DAOPHOT}
\citep{daophot}, i.e., without positional constraints on source
detections. Comparing the results from {\it DAOPHOT} and {\it EM
Photometry} for relatively isolated sources, we find a good agreement
for NUV$<24$, and a gradually increasing difference at fainter
magnitudes, up to 0.21 mag at the catalog limit (DAOPHOT photometry
being brighter). The difference can likely be attributed to unresolved
detections in DAOPHOT, and it also represents the upper limit on the
above-discussed bias introduced by forcing prior extractions. With an
NUV catalog in hand, we repeat the procedure to obtain FUV photometry,
now using NUV$=26.5$ to set the cut on the prior catalog and adopting
a FUV$=26.5$ cut for the final FUV catalog (roughly a 3$\sigma$
limit). We estimate the bias at the faint end to be below 0.13
mag. The FUV and NUV catalogs are matched to the CFHTLS catalog by
construction, which is in turn matched to DEEP2 positions (\S
\ref{ssec:cfhtls}). Of CFHTLS sources matched to the full DEEP2
redshift catalog, 22\% have fluxes in FUV caalog and 59\% in NUV
catalog. RMS calibration errors of 0.052 and 0.026 mag are adopted for
FUV and NUV respectively \citep{morrissey}.

\subsection{CFHT Legacy Survey optical photometry} \label{ssec:cfhtls}

The EGS represents one of the four deep fields targeted by the CFHT
Legacy Survey (CFHTLS). The central region of the EGS is observed with
the MegaPrime/MegaCam imager/detector in a single pointing covering a
$1\degr \times 1\degr$ field of view (Figure \ref{fig:area}) in five
optical bands ($u^*g'r'i'z'$). The limiting magnitudes corresponding
to 80\% completeness are 27.2, 27.5, 27.2, 27.0, 26.0, respectively.
\footnote{http://www1.cadc-ccda.hia-iha.nrc-cnrc.gc.ca/community/CFHTLS-SG/docs/cfhtls.html}
We use band-merged catalogs (publicly available version 2008A) based
on $i$-band detections, with aperture photometry measured from
$i$-band derived apertures. Matching to the DEEP2 redshift catalog is
performed using a $0\farcs3$ search radius. Astrometric zero points
coincide to within $0\farcs02$, and the 1-D coordinate scatter between
the two catalogs is $0\farcs08$, i.e., both catalogs have very
accurate astrometry. There are no multiple matching candidates. Of
9923 DEEP2 objects (from the full redshift catalog not restricted to
good quality redshifts) that lie within CFHTLS coverage, 9056 (91\%)
are matched.  Based on the scatter of the comparison of the bright end
with SDSS, we adopt RMS calibration errors of
(0.04,0.025,0.025,0.025,0.025,0.035) mag for ($u^*g'r'i'z'$).

\subsection{MMT $u$-band photometry}

In addition to $u^*$-band data from CFHTLS, we also use $u'$-band
photometry obtained with MegaCam \citep{mmt} on the MMT. These data
extend across nearly the entire length of the EGS, with 24 overlapping
fields each covering $0.4\degr \times 0.4\degr$. The $5\sigma$
limiting magnitude varies between 26.3 and 27.0. Matching to the
DEEP2 redshift catalog was performed using a $0\farcs4$ search radius
after applying a $0\farcs18$ offset in declination to bring the
coordinate system of MMT data (based on USNO-B1) into agreement with
the SDSS system used in DEEP2. The 1-D coordinate scatter between the
two catalogs is $0\farcs12$. Of 15283 DEEP2 objects (from the full
redshift catalog) within MMT coverage, 10965 have a match (72\%), with a
handful of multiple match candidates, in which case the object with
brighter $u'$ is selected.

Since we have $u$-band photometry from CFHTLS as well, we can compared
them.  The scatter between the MMT and CFHTLS $u$-band magnitudes does
not increase with DEEP2 matching separation, indicating that the
matches are real throughout the search radius. However, there is a
0.08 mag overall offset between two magnitudes in the sense that
CFHTLS $u$ is fainter than MMT $u$. At the bright end we can compare
these magnitudes to SDSS. MMT $u$ matches SDSS very well, while CFHTLS
$u$ is again fainter, but by 0.05 mag (both MMT and CFHTLS photometry
was first transformed to SDSS system). The offsets between MMT and
CFHTLS $u$ do not show an obvious color dependence.  We correct these
offsets in the SED fitting. We adopt a calibration RMS error of 0.04
mag for MMT photometry.

\subsection{Palomar $K$-band photometry}

The reddest photometry band that we use in the SED fitting comes from
the Palomar $K$-band survey of DEEP2
\citep{bundy06,bundy07}. Including redder bands (such as IRAC 3.6 and
4.5 \mic) would not place additional constraints on SFRs, which are
the main focus of this paper. The EGS is almost fully covered with
thirty-five $8\farcm6 \times 8\farcm6$ WIRC frames, down to a
21.7--22.5 mag limit at 80\% completeness. We use MAG\_AUTO fluxes
from the \citet{bundy06} SExtractor catalog, and their $1\farcs1$
matching to DEEP2. Of 16087 DEEP2 objects from the full redshift
catalog, most of which are within $K$ survey coverage, 10398 (65\%)
have a match.

\subsection{MIPS {\it Spitzer} 24 \mic\ photometry} \label{ssec:mips}

In addition to UV through near-IR data that are used for the SED
fitting, we use 24 \mic\ observations to estimate IR luminosities. The
24 \mic\ data were obtained with MIPS on {\it Spitzer} as part of the
FIDEL survey. FIDEL observed EGS and ECDF-S fields with MIPS at 24 and
70 \mic\ to depths of 30 $\mu$Jy and 3 mJy, respectively. These depths
approach those of GOODS yet cover a larger area. In EGS, these data
are five times deeper than the previous data described in
\citet{aegis} (which are co-added to FIDEL data). We extract PSF
fluxes from 24 \mic\ images using {\it DAOPHOT} (MIPS has $6''$
resolution at 24 \mic; \citealt{mips}). We then match 24 \mic\ sources
having S/N$>3$ (corresponding to 10-16 $\mu$Jy) to the CFH12K
photometry catalog \citep{coil} using a $1\farcs5$ matching
circle. This search radius is appropriate for bright 24 \mic\ sources,
which have a 1-D astrometry precision of $0\farcs5$. However, fainter
sources have poorer astrometry, so we subject sources that initially
had no match (39\% of total) to a larger $3''$ radius search,
recovering some 60\% of them. In cases of multiple optical candidates
(4\% of cases), we pick the one that has the $I$-band to 24 \mic\ flux
ratio that is at least two times more likely than that of other
candidates, where the probability is based on the flux ratio
distribution of unique matches. This allows us to resolve 30\% of
multiple matches. We consider the remaining multiple matches to be a
blend of more than one optical source and exclude them from the
catalog of matched sources and from further analysis. Altogether, an
optical match is determined for 74\% of 24 \mic\ sources within the
optical coverage. The unmatched 24 \mic\ sources are either blends or
are presumably fainter than the $R=24.5$ limit of CFH12K photometry
catalog. Similar detection rates (for similar $R$ limits) were found
in CDF-S by \citet{pablo05} and by \citet{lefloch}, 70\% and 60\%,
respectively. In the opposite direction, of DEEP2 objects from the
full redshift catalog, 6581 (41\%) are detected at 24 \mic. We decide
to match 24 \mic\ data directly to optical instead of using IRAC
photometry \citep{barmby} as an intermediate step, because IRAC
coverage of EGS is not as extensive. As a test, for areas with IRAC
coverage we run matching via IRAC and find that in 98\% of cases we
obtain the same optical match as with direct 24 \mic\ to optical
matching.

\subsection{Other data and data products}

In addition to redshifts, DEEP2 spectra provide emission line fluxes
which we use to select narrow-line AGNs. Derivation of fluxes is
described in \citet{weiner}. We also use {\it Chandra} X-ray
detections from AEGIS-X DR2 to select X-ray AGN. Details of the X-ray
data, catalog construction and matching to optical sources are given
in \citet{laird}.

\section{UV/optical SFRs from SED fitting} \label{sec:sed}

The sample used in SED fitting consists of DEEP2 galaxies with secure
redshifts and spectra classified as galaxies.  A small number of
galaxies fitting an AGN template (broad-line AGNs, QSOs) are excluded
since their continua will be affected by the light from the active
nucleus, and therefore cannot be fitted with our models. In terms of
area, our sample lies in the overlap of CFHTLS and \galex\ regions
(dark gray region in Figure \ref{fig:area}), which contains 5878 DEEP2
galaxies. Other data cover this region fully, so their footprints are
not relevant. Using a technique similar to that of \citet{blanton} we
first estimate the area of the full DEEP2 EGS (light and dark gray
region in Figure \ref{fig:area}). Our sample contains 53\% of the
total number of sources in full area, from which we arrive at an
estimate of 0.31 sq.\ deg.\ for the overlap area (dark gray region in
Figure \ref{fig:area} . We remind the reader that detections are not
required in all bands as long as the object comes from the overlap
area, so our {\it optical} sample used in SED fitting remains only
$R$-band selected.

We estimate galaxy parameters such as the star formation rate, dust
attenuation, stellar mass, age, rest-frame colors and magnitudes,
using the stellar population synthesis models of
\citet{bc03}.\footnote{An update of \citet{bc03} models is being
developed to address issues concerning the treatment of TP-AGB stars
\citep{maraston,bruzual}. However, these changes will have almost no
effect on SFRs. The {\it systematic} effect on the stellar masses will
also be limited since we do not use the IRAC bands.} The methodology
is basically identical to that used in \citet[S07]{s07}, and we refer
the reader to that paper for details of stellar population and
attenuation models. Model libraries are built by considering a wide
range of star formation histories (exponentially declining continuous
SF with random stochastic bursts superimposed), with a range of
metallicities (exact ranges are given in S07). Each model is
dust-attenuated to some degree according to a two-component
prescription of \citet{cf00}. This model assumes that young
populations ($<10$ Myr) lie within dense birth clouds and experience
total optical depth of $\tau_V$. When these clouds disperse, the
remaining attenuation is only due to the general ISM, having optical
depth of $\mu\tau_V$, where $\mu$ is typically $\sim 0.3$. In both
cases the extinction law of a single population is $\propto
\lambda^{-0.7}$. In our models, we allow for a range of $tau_V$ and
$\mu$ values as described in S07. A feature in the SED that has the
greatest weight in constraining the dust attenuation is the UV slope,
which is steeper (the UV color is redder) when dust attenuation is
higher \citep{calzetti94}. However, there is a significant scatter
between the UV slope and the dust attenuation due to the differences
in the SF history \citep{kong}, which in our model is constrained by
the inclusion of optical data. Finally, we include reddening due to
the intergalactic medium, according to \citet{madau}. SFRs and stellar
masses are determined assuming a Chabrier IMF.

The only difference in model libraries with respect to S07 is that we
now construct them in $0\leq z_{\rm lib} \leq 1.6$ range, at 0.1
intervals in $z_{\rm lib}$, whereas in S07 libraries extended out to
$z_{\rm lib}=0.25$ at 0.05 intervals. While library redshift
resolution is finite, note that we use exact galaxy redshifts to scale
mass and SFR from normalized model quantities to full absolute
values. We test the effects of library redshift coarseness on the
derivation of SFR and mass. On average, redshift and $z_{\rm lib}$
differ by 1/4. We produced a test run where we increase this
difference to 3/4 by assigning the next or the preceding library
(e.g., galaxy at $z=0.97$ is fitted with $z_{\rm lib}=1.1$ models
instead of $z_{\rm lib}=1.0$). As expected, the average values of SFR
and stellar mass do not change, but the average absolute difference is
0.13 dex for SFR and 0.09 dex for stellar mass. From this we can
extrapolate that when redshift and $z_{\rm lib}$ differ by 1/4 this
deviation will be 0.04 and 0.03 dex, respectively. In our analysis
this will be reflected as the small addition to the random
errors. Finally, since only models with formation age shorter than the
age of the universe at $z_{\rm lib}$ are allowed, the number of model
galaxies decreases from $10^5$ at $z_{\rm lib}=0$ to $3\times 10^4$ at
$z_{\rm lib}=1.6$. Even at the high-redshift end the number of models
is sufficiently large not to introduce biases in the derived
parameters \citep{s07}.

Our SED fitting involves up to 9 flux points (FUV, NUV, $u'_{\rm
MMT}$, ($u^*g'r'i'z')_{\rm CFHTLS}$, $K$), their photometric errors,
and the redshift. Photometry for various bands has been derived in a
heterogeneous manner, but it should reflect the total fluxes in most
bands. This will have a negligible effect on the results. The SED
fitting has one degree of freedom (scaling between the observed and
the model flux zero points). For each galaxy the observed flux points
are compared to model flux points, and the goodness of the fit
($\chi^2$) determines the probability weight for the given model, and
thus of the associated model parameters in the final probability
distribution function (PDF) of each parameter (such as the SFR,
stellar mass, etc.). We then use the average of the probability
distribution as our nominal estimate of a galaxy parameter and
consider the width of the probability distribution function as an
estimate of parameter error and its confidence range. In cases where
no detection is present in a given band, that band does not contribute
to $\chi^2$. The Bayesian SED fitting performed here has many
advantages with respect to more traditional maximum likelihood
method. The parameter PDFs allow us to determine how well a given
parameter can be determined taking into account not only the
observational errors, but also the degeneracies among the models. For
example, suppose that the dust attenuation and the metallicity were
completely degenerate, i.e., that various combinations of the two
produce identical SEDs. While the maximum likelihood will pick one
(basically arbitrary) SED and its parameters as the best fitting, the
Bayesian fitting will produce a wide flat PDF suggesting that many
different values are equally probable. Similarly, the lack of
observational constraints will also be reflected in the increased
width of PDFs of those parameters that rely on these
observations. While all flux points contribute to all galaxy
parameters, it is to be expected that the UV fluxes will be more
critical in obtaining current SFRs and dust extinctions, while flux
points red-ward of 4000\AA\ will contribute more to the determination
of the stellar mass. Also, we note that despite the fact that our
input (observed SED) contains limited information content, one could
in principle derive an unlimited number of galaxy parameters, since
the PDF of each parameter will correctly marginalize over
observational and model uncertainties. Most of these parameters will,
of course, not be independent, which one can again establish using
(multidimensional) PDFs. Reader is referred to S07 for further details
about the SED fitting procedure. \citet{walcher}, who use very similar
model libraries and the fitting technique, also provide extensive
discussion on the method and its robustness (their \S 2). In \S\
\ref{sec:lir} we will discuss in more detail the errors in the SFR.

Before performing the SED fitting, we first examined color-color
diagrams where we plot observed colors in some redshift interval
together with model colors corresponding to that redshift. We were
prompted to perform these tests after learning that there could be a
discrepancy between the observed and \citet{bc03} colors in the VVDS
sample, in the sense that models were underestimating the flux in the
3300-4000 \AA\ range \citep{walcher}. By visually comparing the locus
of observed and model colors for various combinations of colors and
redshift bins, we were mostly able to confirm this effect. We find the
level of discrepancy (up to 0.2 mag) to be similar for both blue and
red galaxies, which makes it unlikely to be the result of
contamination from [OII]$\lambda$3727 emission line (emission lines
are not included in \citealt{bc03} modeling) but, rather caused by
differences in the continuum. \citet{walcher} use an updated version
of \citet{bc03} models (which include new prescriptions for TP-AGB
stars) and still encounter the discrepancy. However, both the original
\citet{bc03} models used here and the updated version used by
\citet{walcher} are based on same {\it stellar libraries} which have a
transition from empirical STELIB spectra to synthetic BaSeL spectra at
3200 \AA.  It is beyond the scope of this work to try to understand
the origin of this problem. Since at a given redshift this discrepancy
would affect only one of our flux points, we decide to exclude that
flux point from the SED fitting, i.e., we exclude $g$ at $0.3\leq
z_{\rm lib}\leq 0.5$, $r$ at $0.7\leq z_{\rm lib}\leq 0.9$, $i$ at
$1.0\leq z_{\rm lib}\leq 1.3$, and $z$ at $z_{\rm lib}=1.4$.

We require a minimum of three flux points for the SED fitting, though
most galaxies have many more. In 336 cases this criterion is not met
(mostly because CFHTLS magnitudes are not measured in spite of the
fact that the object is listed), and we exclude these objects from
further analysis. Additionally, we eliminate 197 objects with poor SED
fits (i.e., high $\chi^2$) whose galaxy parameters are unreliable.  In
S07 we discuss galaxies with bad SED fits and conclude that they
mostly result from bad data rather than the limited parameter space of
the models. Thus we arrive at the final {\it optical} sample of 5345
objects for which we obtain galaxy parameters from the SED fitting.

\begin{figure}
\epsscale{1.2}
\plotone{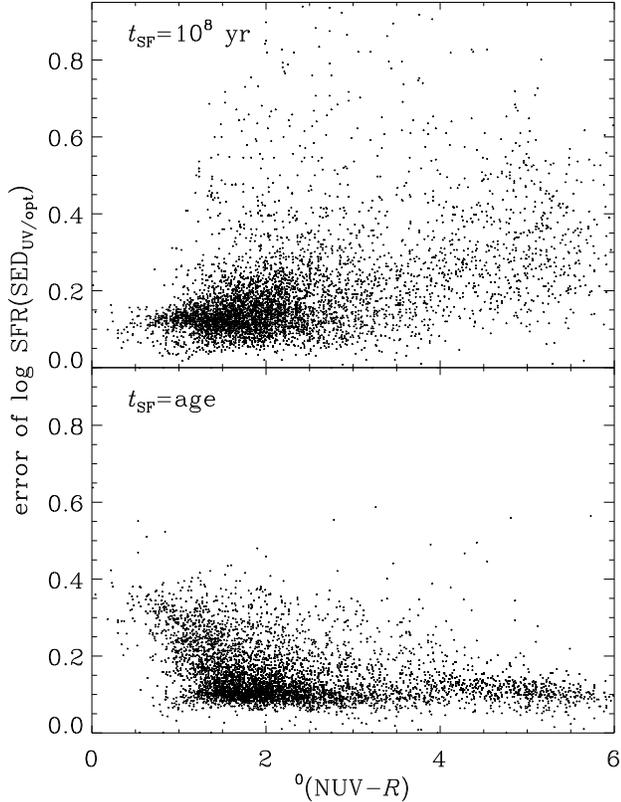}
\caption{Errors in the current (upper panel) and age-averaged (lower
panel) star formation rates, both corrected for dust attenuation and
plotted against the rest-frame color. Errors are estimated from the
width of the probability distribution function and take into account
measurement and model uncertainties, such as the uncertainty in the
dust correction.}
\label{fig:sfrerr}
\end{figure}

Two parameters derived from the UV/optical SED fitting will feature
most prominently in this work: the ``current'' SFR (i.e., the SFR
averaged over the last $10^8$ yr, the shortest timescale that can be
reliably probed with stellar continuum) and the ``age-averaged'' SFR
(i.e., the SFR averaged over the dominant population age, which
depends on a galaxy and typically ranges between 1-3 Gyr). Both will
be discussed in more detail later. Here we wish to assess the typical
errors associated with these parameters. Both SFRs are dust-corrected
and their probability distribution functions will automatically
reflect various sources of uncertainty. In the case of the current SFR
the error will be dominated by the uncertainties in the dust
correction, which we confirm by finding a strong correlation between
SFR error and dust correction error. In Figure \ref{fig:sfrerr} (upper
panel) we show the error in current SFR ($t_{\rm SF}=10^8$ yr) as a
function of rest-frame ${\rm NUV}-R$ color, which we will use to
select actively star-forming (blue) and quiescent (red) galaxies. The
error equals 1/4 of the 95\% confidence range of a PDF, which in the
case of a Gaussian distribution would correspond to a standard
deviation ($1\sigma$). The majority of galaxies have errors below 0.2
dex (60\%). As expected, the errors increase as one moves towards
redder, less actively star-forming galaxies. Some galaxies, regardless
of color, have an error in excess of a factor of 3 (0.5 dex). We find
that these galaxies are very faint in the UV, with rest-frame FUV
magnitudes fainter than 24.5. Figure \ref{fig:sfrerr} (lower panel)
shows errors in age-averaged SFR. Unlike the current SFR which is
mostly constrained by rest-frame UV, the age-averaged SFR is
determined by optical light of stars having ages 1-3 Gyr. It is also
typically less than 0.2 dex, but it is on average higher for blue
galaxies where the optical light is fainter. Unlike the current SFR,
here the errors stay below 0.5 dex, owing to the lack of very faint
optical sources (the sample being $R$-band selected). While the
stellar mass doesn't figure prominently by itself in this work, let us
mention that the typical stellar mass errors are below 0.1 dex.

\begin{figure}
\epsscale{1.2}
\plotone{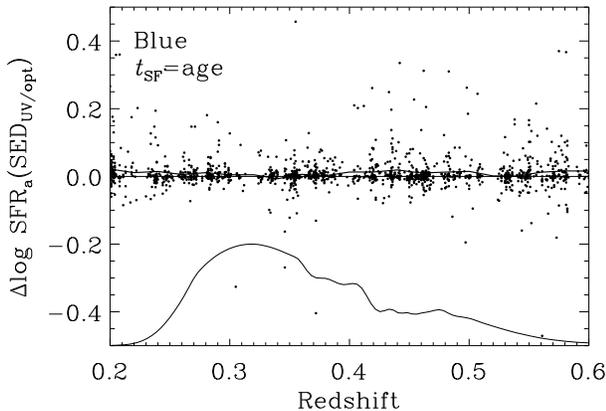}
\caption{The effect of H$\alpha$ contamination on the SFRs derived
from the SED fitting. In the displayed redshift range the H$\alpha$
passes through $z$ filter (its relative contribution to $z$ is shown
as a solid curve with arbitrary amplitude) and could thus affect the
SED fitting. Shown is the difference in age-averaged SFRs (SFR with
subscript 'a') when the $z$ band is excluded from the fitting,
compared to nominal fitting. Mean residuals (solid line slightly above
the $y=0$ line) are below 0.02 dex and are not correlated with the
expected H$\alpha$ contribution. We plot only blue, star-forming
galaxies for which the contribution of H$\alpha$ should be the
largest.}
\label{fig:dsfrt_noz}
\end{figure}

Galaxy SED models used in this work come from stellar population
synthesis alone, without the inclusion of gas emission lines. This
could potentially lead to systematic effects in the parameters derived
from the SED fitting, since the emission lines could ``contaminate''
the broad-band fluxes used in the fitting. Typically, the most
luminous emission line in our sample is H$\alpha$, followed by
[OII]$\lambda$3727. H$\alpha$ becomes redshifted beyond the reddest
optical band ($z$ band) at redshifts above 0.4, so it does not affect
most of the galaxies in our sample. For those at lower redshift we
estimate the effect of H$\alpha$ contribution by running the SED
fitting without the $z$ band and comparing the resulting SFRs to those
from the nominal fitting. The residuals of age-averaged SFRs are shown
against the redshift in Figure \ref{fig:dsfrt_noz} for blue (mostly
star-forming) galaxies. We present age-averaged SFRs since they should
be more affected by the $z$-band flux than the current SFRs. Typical
average residuals are below 0.02 dex, and there is no obvious
correlation with the expected relative contribution of H$\alpha$ to
$z$-band (solid curve, shown with arbitrary amplitude). Residuals tend
to be positive, which actually corresponds to SFRs from the fit
without the $z$ band being larger than the nominal ones, the opposite
from what is expected if H$\alpha$ raises the $z$-band flux. As for
[OII] line, one cannot evaluate its potential effect on broad-band
fluxes because of the unrelated issues with models in the 3300-4000
\AA\ range (discussed previously in this section). Since we already
exclude from SED fitting the bands that sample this wavelength range,
any effects of [OII] will be removed from our nominal results.

\section{Infrared and optical properties of 24 \mic\ sample} \label{sec:lir}

\begin{figure}
\epsscale{1.2}
\plotone{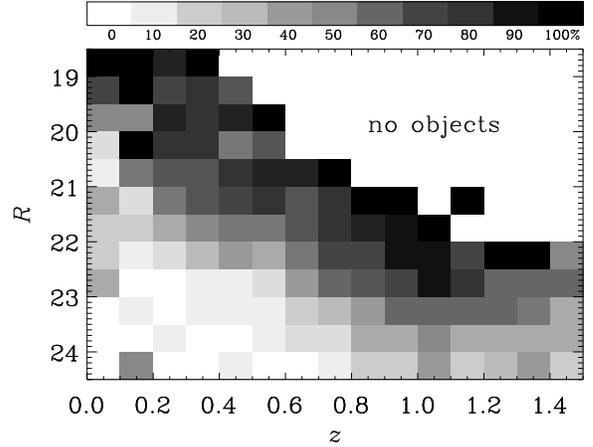}
\caption{Detection fraction of MIPS 24 \mic\ observations in bins of
redshift and apparent magnitude. Gray pixels represent the detection
fraction, with black being 100\% and white being zero, except in the
upper right corner where there are no objects.}
\label{fig:r_z_hist}
\end{figure}

\begin{figure}
\epsscale{1.2}
\plotone{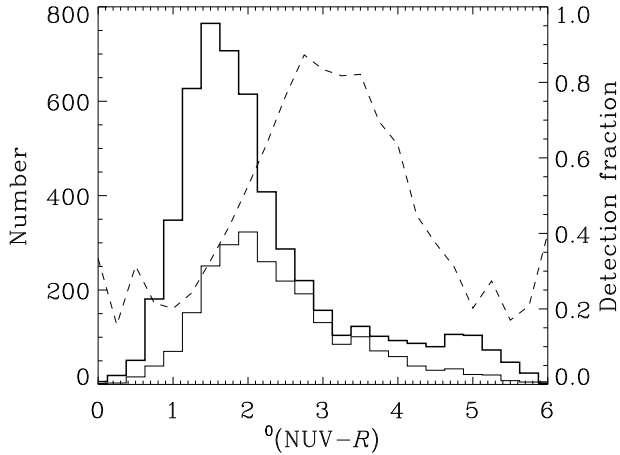}
\caption{Rest-frame color distribution of the sample. Bold histogram
shows the distribution of our optical ($R<24.1$) sample (for which the
SED fitting is performed), and the thin histogram shows the
distribution of sample sources which are detected at 24 \mic. The
dashed line is the ratio of the two histograms, i.e., 24 \mic\
detection fraction. The 24 \mic\ detection fraction peaks at the red
end of the blue sequence and in the ``green valley''. In this and
subsequent figures, superscript zero in NUV$-R$ designates rest-frame
(note that NUV$-R$ is not dust corrected).}
\label{fig:nuvr_hist}
\end{figure}

Out of 5345 objects in the optical sample, we have 24 \mic\ detections
for 2430 (45\%). The 24 \mic\ imaging covers 99.5\% of the area of the
optical sample, with exposure times varying across the field from
$\sim0$ to 19 ks (average exposure is 10 ks). The optical source
detection efficiency grows linearly with the logarithm of the exposure
time; it is 26\% at 1 ks, and 56\% at 19 ks. In Figure
\ref{fig:r_z_hist} we plot the 24 \mic\ detection efficiency as a
function of redshift and apparent $R$ magnitude. The gray scale is
proportional to the detection fraction, with black representing 100\%
and white being zero, except at bright magnitudes and high redshifts
where there are no objects. At each redshift the detection fraction
increases with optical brightness, but for a given $R$ magnitude the
efficiency increases with redshift, which is the consequence of the
detection efficiency rising with absolute magnitude. In Figure
\ref{fig:nuvr_hist} we plot the distribution of rest-frame ${\rm
NUV}-R$ colors determined from the SED fitting. The optical sample
(bold histogram) is dominated by galaxies lying in the blue sequence
($^0({\rm NUV}-R)<3.5$)\footnote{Throughout the text, superscript zero
designates rest-frame and not that the color is dust-corrected.}. The
peak of the red sequence ($^0({\rm NUV}-R)>4.5$) is less obvious
because the $R$-band selection eliminates fainter red galaxies at
higher redshifts. The thin-line histogram shows galaxies with 24 \mic\
detection. Again, most detections are of blue galaxies. The ratio of
the two histograms represents the 24 \mic\ detection fraction and is
plotted as the dashed curve with a corresponding axis on the
right-hand side. The 24 \mic\ detection efficiency strongly peaks at
intermediate colors, including the so called ``green valley''
($3.5<^0({\rm NUV}-R)<4.5$), where it reaches $>80\%$. A similar
result was recently obtained by \citet{cowie}.

\begin{figure}
\epsscale{1.3}
\plotone{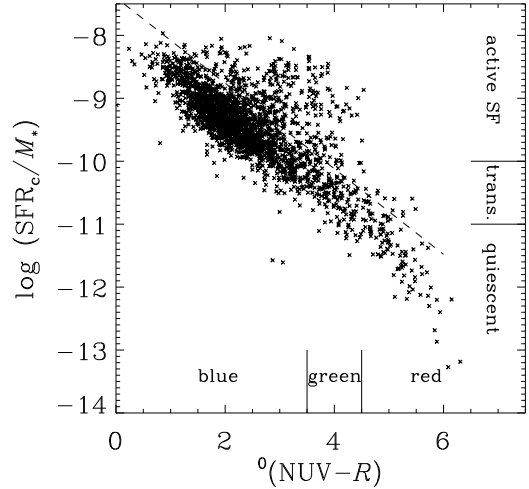}
\caption{Relation between the rest-frame UV-to-optical color and the
specific SFR of the part of the optical sample detected at 24
\mic. The two quantities are roughly equivalent, except that the
specific SFR, derived in UV/optical SED fitting, is
dust-corrected. Also, SFR is averaged over $10^8$ yr, which we call
``current'' SFR (subscript 'c'). In cases of moderate dust
UV-to-optical color correlates with the SF history of a galaxy (the
trend of the majority of galaxies in this sample). Galaxies that
scatter away from that relation are dusty starbursts (objects above
the dashed line). Note that many genuine red sequence (and not just
reddened), i.e. quiescent, galaxies are still detected at 24 \mic. The
classification of colors into blue, green, and red, and the SF
histories into active star-forming, transitional and quiescent is
based on local ($z<0.2$) studies that employ photometry and
spectroscopy to distinguish between these categories
\citep{wyder,s07}.}
\label{fig:ssfr_nuvr}
\end{figure}

Motivation for the above division into blue and red sequence galaxies
comes from a marked bimodality in optical colors of local ($z<0.2$)
galaxies \citep{strateva}, where blue sequence galaxies have active
star formation, while red have generally ceased forming stars. The
introduction of UV-to-optical colors by {\it GALEX} led to a
recognition of a region in between the blue and the red sequences that
was not prominent in optical colors \citep{wyder}. Galaxies that
occupy this region, the green valley, acquire intermediate colors
either because they have an intermediate SF history (transitional
galaxies), or because their colors have been reddened by dust and
would otherwise be blue (dusty starbursts) \citep{martin,s07}. One can
distinguish between the two by plotting the specific SFR (SFR/$M_*$)
against the color. This is shown in Figure \ref{fig:ssfr_nuvr}, where
the rest-frame color, dust-corrected SFR and the stellar mass come
from our UV/optical SED fitting, and the SFR is what we call current,
i.e., averaged over $10^8$ yr. If dust reddening is moderate, there
should be a correlation between the specific SFR (basically a ratio of
recent to past SF) and the rest-frame UV-to-optical color
\citep{s05}. One can see that this is the case for the majority of
galaxies, especially in blue and red regions. However, if a galaxy has
{\it dusty} star formation, it will have redder colors for a given
specific SFR (because the SFR is corrected for effects of dust, while
color is not). Such galaxies scatter above the main trend in Figure
\ref{fig:ssfr_nuvr} (galaxies above the dashed line). In terms of
colors, dusty starbursts are present in the blue sequence and the
green valley, with the relative number of dusty to non-dusty systems
peaking in the green valley. Here we note that even after accounting
for dust attenuation there still exist 24 \mic\ detections among
fairly quiescent galaxies ($\log ({\rm SFR}/M_*)<-11$). The source of
their mid-IR emission will be discussed in \S \ref{sec:lir_red}.

\begin{figure}
\epsscale{1.2}
\plotone{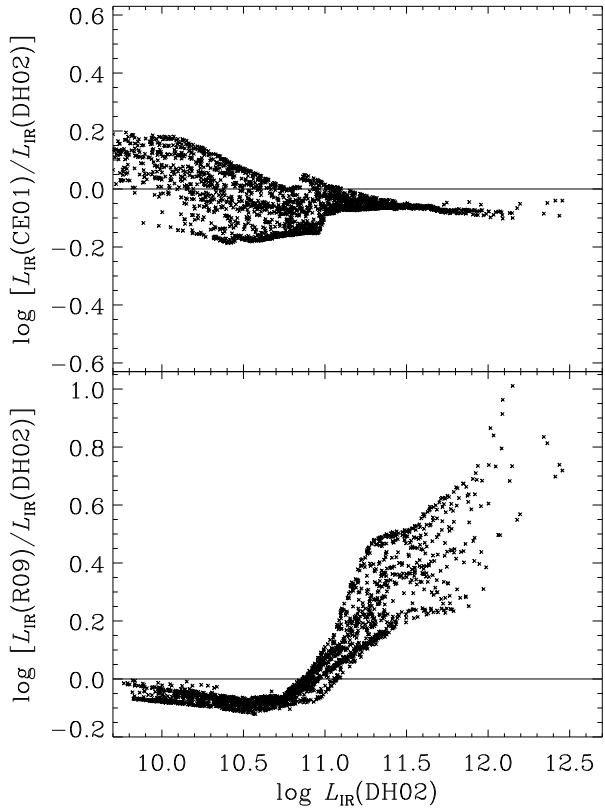}
\caption{Comparison of IR luminosities obtained with various IR
templates and the 24 \mic\ flux point. Comparison is against $\lir$
obtained with \citet{dh02} templates (DH02). The difference is
relatively small with respect to \citet{ce01} templates (CE01), but
more significant with respect to \citet{rieke} templates (R09),
especially at higher luminosities. We use IR luminosities based on
\citet{dh02} in the rest of the paper.}
\label{fig:dlir_comp}
\end{figure}

To infer the total infrared luminosity, $\lir$(8--1000\mic), we fit 24
\mic\ flux densities and redshifts to infrared SED templates of
\citet{dh02}.\footnote{In most of the paper {\it total} infrared
luminosity will simply be called ``infrared luminosity'', or $\lir$.}
These templates were normalized to follow the local {\it
IRAS}-calibrated far-IR color vs.\ luminosity relation of
\citet{marcillac}. The assumption is that for most galaxies in our
sample the mid-IR flux is representative of the total IR luminosity,
and that one can use the luminosity-dependent SED models to constrain
it. This is certainly an oversimplification. The rest-frame wavelength
range probed by our 24\mic\ observations (10--18\mic) contains many
PAH lines whose relation to the mid-IR continuum and to the total IR
luminosity may vary significantly compared to the fixed ratios assumed
in the templates \citep{jdsmith}. Also, translating mid-IR
luminosities to total ones introduces potentially large uncertainties
in the $K$ correction. However, the use of total vs. mid-IR continuum
luminosities is not critical in this work, and, as we will show in
\S\ref{ssec:mono}, the results do not change if we use monochromatic
mid-IR luminosity instead. The use of total IR luminosity is motivated
by the commonality with which this measure is interpreted as a star
formation rate, especially in high-redshift studies. Once we obtain
the interpolated IR template, we calculate the total infrared
luminosity according to the relation of \citet{sanders} (directly
integrating the SED produces very similar results, and Sanders \&
Mirabel definition is used as a convention). In addition to $\lir$
derived from \citet{dh02} templates, we additionally calculate $\lir$
based on luminosity-dependent templates of \citet{ce01} and recent
templates of \citet{rieke}. The comparison of the two with respect to
$\lir$ from \citet{dh02} is shown in Figure \ref{fig:dlir_comp}. For
our sample the IR luminosities from \citet{ce01} and \citet{dh02} stay
within 0.2 dex of one another (average difference is -0.03 dex and the
standard deviation of the ratio is 0.09 dex). At $\log \lir
\rm{(DH02)}>11$ the scatter in the ratio is very small (0.02 dex), and
the difference is almost constantly around -0.06 dex (\citet{dh02}
estimate being higher). Differences with respect to \citet{rieke} IR
luminosities are much higher, especially for LIRGs and ULIRGs ($\log
\lir \rm{(DH02)}>11$). \citet{rieke} estimates get increasingly
discrepant as the luminosity increases (up to an order of magnitude),
to the extent that 16\% of what are classified as LIRGs according to
\citet{dh02} become ULIRGs with \citet{rieke} templates, while the
number of ULIRGs changes from 21 to 156. Using \citet{rieke} templates
could possibly affect some of the results in our work, but also those
of many other studies. On the other hand, the differences between the
other two templates are smaller (see also \citealt{pablo}), and we
adopt IR luminosities based on \citet{dh02} templates as our nominal
values. All of these templates are based on local, star-forming
galaxies, so they may not be entirely appropriate for high redshift
galaxies such as those in our sample, or to more quiescent
galaxies. In our analysis we will therefore use caution when
interpreting the IR luminosities.

\begin{figure}
\epsscale{1.2}
\plotone{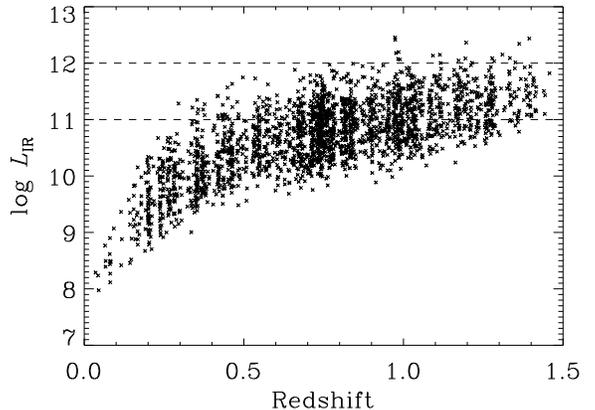}
\caption{Infrared luminosity $\lir$, derived from 24 \mic\ flux, as a
function of redshift. Two dashed lines define LIRGs ($11<\log
\lir<12$), and ULIRGs ($\log \lir>12$). Normal star forming galaxies
($\log \lir\sim10$) are detected to $z\sim0.7$. There are very few
ULIRGs in our sample.}
\label{fig:lir_z}
\end{figure}

In Figure \ref{fig:lir_z} $\lir$ is shown as a function of
redshift. Two dashed lines show regions that define LIRGs and
ULIRGs. LIRGs start to dominate raw counts at $z\sim0.8$ and we remain
sensitive to LIRG luminosities out to the upper redshift limit. We are
also sensitive to normal star forming galaxies ($\lir \sim 10^{10}
L_{\odot}$) to $z\sim0.7$. The number of ULIRGs is small even at the
highest redshifts. This is similar to the luminosity distribution
presented in \citet{lefloch} for CDF-S. Since we study only 24 \mic\
detections with available spectroscopic redshifts, we check if the
redshift selection introduces any biases at the high-luminosity end of
$\lir$. For this purpose we consult a catalog of {\it photometric}
redshifts based on CFHTLS photometry \citep{ilbert}, and match it to
optical counterparts of 24 \mic\ sources. We then compute $\lir$ based
on photometric redshift. For $\lir>10^{9.4} L_{\odot}$ the
distribution of $\lir$ of the photometric redshift sample matches the
shape of the distribution of $\lir$ in our spectroscopic redshift
sample, implying no bias of the latter at high IR luminosities.

\section{Infrared luminosity and UV/optical SFR in blue sequence
galaxies} \label{sec:lir_sfr}

We now have on one hand dust-corrected SFRs constrained from
UV/optical SED fitting, and on the other hand infrared luminosities
from 24 \mic\ flux. We will often refer to dust-corrected UV/optical
SFRs as SED SFRs, or just SFRs. We emphasize that SED fitting allows
us to construct SFRs on various timescales, i.e., SFR averaged over
some time interval. They are chosen to be $t_{\rm SF}=10^7$, $10^8$,
$10^9$ and $2\times10^9$ yr (averaging interval $t_{\rm SF}$ ends with
the epoch of the observation, i.e., it is not centered on it). In
addition to these fixed timescales, we also determine SFR averaged
over the age of the galaxy, which is calculated as the total stellar
mass (current mass plus the recycled mass as estimated in our models)
divided by the time since galaxy formation (from models). Because in
our models the galaxy formation ages have a {\it uniform}
distribution, i.e., they are not restricted to some high-redshift
galaxy formation epoch, the derived formation age will be largely
driven by the age of the {\it dominant population} in terms of light
production (for example, blue galaxies will be assigned young
``formation'' ages regardless of their ``real'' age). We verify that
there is a very tight correlation between the derived ``formation''
age and the light-weighted age. Therefore, what is actually measured
by age-averaged SFR is the average SFR over the age of the dominant
population, which for blue-sequence galaxies in our sample varies
between 0.1 and 3 Gyr. Errors in UV/optical SFRs were discussed in \S
\ref{sec:sed} for the full optical sample and those conclusions are
applicable here for the subset detected at 24 \mic.

Since the IR luminosity is usually considered in the context of
(current) star formation rate, we begin our analysis using the concept
of star formation rate, but extending it to include SFRs averaged over
longer time periods. The temporal aspect is not essential here. The
timescales used in averaging the SFR simply correspond to the light
emitted {\it today} by stellar populations of different {\it
ages}. Therefore, what constrains the SFRs averaged over progressively
longer timescales is the rest-frame luminosity at increasingly redder
wavelengths. We will return to this relation between SFR and
luminosity later. 

\begin{figure}
\epsscale{1.3}
\plotone{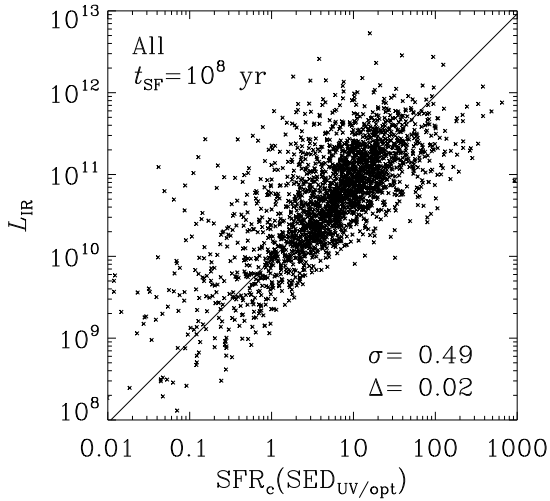}
\caption{Comparison of infrared luminosities and dust-corrected
UV/optical star formation rates for all galaxies. Throughout the paper
the SFRs are given in $M_{\odot} {\rm yr}^{-1}$ and for Chabrier
IMF. UV/optical SFR is averaged over $10^8$ yr (``current'' SFR,
subscript 'c'), the characteristic UV timescale. Numbers in lower
right corner show the dispersion ($\sigma$) and the mean offset
($\Delta$) (in dex) with respect to the 1:1 correspondence between the
SFR and infrared luminosity (solid line) that assumes the
\citet{kennicutt} conversion (converted to Chabrier IMF). The
Kennicutt conversion is derived for dusty star-forming galaxies with
constant SFRs over $10^7$--$10^8$ yr.}
\label{fig:lir_sfr_all}
\end{figure}

The timescale corresponding to lifetime of stars producing the
majority of non-ionizing UV radiation is $\lesssim 10^8$ yr
\citep{kennicutt}. Since young stars are typically assumed to dominate
the dust heating at mid-IR (\S \ref{sec:intro}), we begin by comparing
the infrared luminosity with SED SFR averaged over $t_{\rm SF}=10^8$
yr, which can be regarded a ``current'' SFR. In Figure
\ref{fig:lir_sfr_all} we plot all 2430 galaxies from our optical
sample detected at 24 \mic. The line represents the 1:1 correspondence
between the SFR and the infrared luminosity assuming the Kennicutt
conversion (converted to Chabrier IMF by applying a factor of 1.58,
S07). It is important to recall that the Kennicutt conversion applies
to optically thick dusty starbursts with constant SF histories over
$10^7$--$10^8$ yr and solar metallicities. The conversion factor is
not empirical, but is derived from population synthesis models. While
it is not strictly appropriate to use this conversion for other types
of galaxies (less dusty or less active), such practice is often
encountered. This can be somewhat justified because, as shown in
modeling of \citet{inoue}, the fortuitous cancellation of smaller dust
opacity and the increased IR cirrus causes the Kennicutt conversion to
also hold for less bursty (more normal) SF galaxies. In any case,
conclusions in our work are independent of the validity of conversions
of IR luminosity to SFR, and instead we deal with IR luminosities
directly. In Figure \ref{fig:lir_sfr_all} one sees a good overall
agreement between the IR luminosity and the UV/optical SFR. The points
on average lie 0.02 dex from the 1:1 line. The standard deviation in
the logarithm of $\lir$ to SFR ratio, i.e., the scatter around the 1:1
line, is 0.49 dex (a factor of three). It is the {\it scatter} that we
will use as an indicator of the level of agreement between the IR
luminosity and the SED-derived UV/optical SFR.

\begin{figure}
\epsscale{1.3}
\plotone{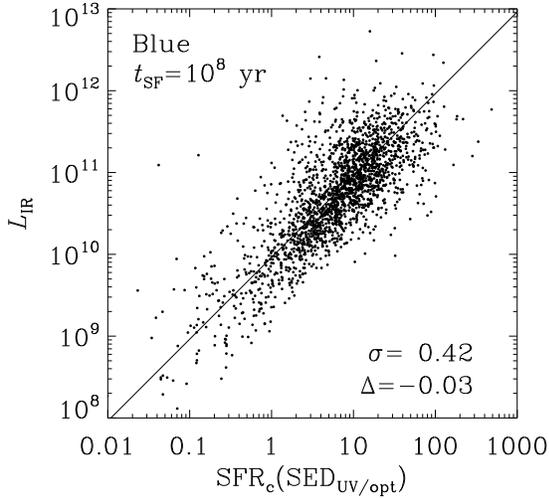}
\caption{Comparison of infrared luminosities and dust-corrected
UV/optical star formation rates for blue-sequence galaxies. Timescale
for SF is still $10^8$ yr, and the \citet{kennicutt} conversion
(derived for dusty star-forming galaxies with constant SFRs over
$10^7$--$10^8$ yr) is shown as the solid line. Removal of red galaxies
reduces the scatter in correlation. Numbers have the same meaning as
in Fig.\ \ref{fig:lir_sfr_all}}
\label{fig:lir_sfr_blue}
\end{figure}

There are a number of potential causes for the level of scatter seen
in Figure \ref{fig:lir_sfr_all}. First, there are measurement
errors. On average, the error in ``current'' SFR is 0.19 dex.  The
average error in 24 \mic\ flux measurement is 6\% (0.02 dex), which is
negligible in comparison. We expect some error from the extrapolation
of the observed 24 \mic\ (rest frame 10--18 \mic) luminosity to total
IR luminosity (e.g, \citealt{lefloch} claim a factor of three
error). As discussed, the two frequently used sets of IR SED templates
already produce a scatter of 0.1 dex in their estimate of
$\lir$. Another source of scatter could be from from the inclusion of
all galaxies in our sample, including many red galaxies with older
stellar populations and not much current star formation. In order to
compare UV/optical SFR and $\lir$ with an assumption that IR arises
from star formation, one needs to limit the sample to actively star
forming galaxies. Following discussion in \S \ref{sec:lir} it would be
appropriate to base such selection on a dust-corrected quantity such
as the specific SFR. However, it is more intuitive to use rest-frame
color instead. Taking blue galaxies ($^0({\rm NUV}-R)<3.5$) will
select most actively star-forming galaxies (including a large number
of dusty starbursts), while not allowing galaxies with more quiescent
SF histories (Figure \ref{fig:ssfr_nuvr}). In Figure
\ref{fig:lir_sfr_blue}, we now compare $\lir$ and SED SFR of blue
galaxies alone. The SFR averaging timescale is still $10^8$ yr. The
scatter in $\lir$ to SFR ratio is 0.42 dex, or 16\% smaller than in
the full sample. The scatter was reduced by the removal of red
galaxies.  This reduction cannot be attributed to slightly smaller SED
SFR errors: 0.17 dex for blue galaxies vs.\ 0.19 dex for the full
sample.

\begin{figure}
\epsscale{1.3}
\plotone{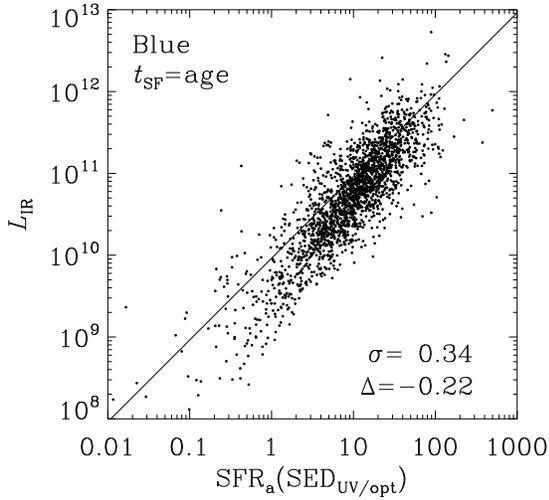}
\caption{Comparison of infrared luminosities and dust-corrected
UV/optical star formation rates for blue-sequence galaxies, where SFR
is now averaged over the galaxy age (i.e., the age of the dominant
population, 0.1--3 Gyr old, subscript 'a'). The correlation between
$\lir$ and SFR is better (the scatter is smaller) than in Figure
\ref{fig:lir_sfr_blue}, where SFR was averaged over $10^8$ yr. The
error in the SFR determination is similar here as it is for SFR
averaged over $10^8$ yr ($\sim0.2$ dex), and does not dominate the
scatter with respect to $\lir$. There is a departure with respect to
\citet{kennicutt} conversion which is derived for dusty star-forming
galaxies with constant SFRs over $10^7$--$10^8$ yr, because most
galaxies have declining SF histories so the SFRs averaged over longer
timescales are on average higher than the SFRs averaged over $10^8$
yr. Numbers have the same meaning as in Fig.\ \ref{fig:lir_sfr_all}}
\label{fig:lir_sfr_t}
\end{figure}

Throughout the LIRG range of luminosities the agreement between the IR
luminosity and UV/optical SFRs is relatively good, albeit with large
scatter. This implies that the LIRGs in the redshift range studied
here cannot be optically thick at UV and optical wavelengths, thus
allowing us to use the stellar continuum to deduce the SFR and other
parameters, such as the stellar mass. This, of course, depends on our
ability to obtain reliable rest-frame luminosities and dust
attenuation estimates. One also sees that the slope between IR
luminosity and UV/optical SFRs is steeper than the \citet{kennicutt}
relation. This is fully expected. The Kennicutt relation applies to
galaxies in which a large fraction of stellar emission is absorbed by
dust. This will be less the case for galaxies with smaller SFRs, which
have smaller dust attenuations \citep{wang}.

Next we explore how the overall scatter in $\lir$ to UV/optical SFR
ratio changes if SFR is averaged over timescales other than $t_{\rm
SF}=10^8$, still restricting our focus to blue galaxies. We again
emphasize that averaging SFR over shorter or longer time periods is a
way to probe the connection of {\it today's} stars having different
range of ages using the SFR concept, and does not imply that the past
episodes of star formation directly affect the IR luminosity that we
see today. We start from $t_{\rm SF}=10^7$ yr, which can be considered
an ``instantaneous'' SFR, and find scatter to be 0.43 dex, slightly
larger than for $t_{\rm SF}=10^8$. Note that UV/optical SED fitting
does not constrain the SFR averaged over such short timescales very
well, and the increase in scatter is simply the result of the poorer
quality of the SFR measure over this timescale. But for the two longer
timescales, $t_{\rm SF}=10^9$ yr and $t_{\rm SF}=2\times10^9$ yr, the
scatter {\it decreases}, to 0.39 and 0.37 dex, respectively. We obtain
yet smaller scatter, 0.34 dex, when we consider SFR averaged over the
age of the dominant population in the galaxy, shown in Figure
\ref{fig:lir_sfr_t}. \notetoeditor{Figures \ref{fig:lir_sfr_blue} and
\ref{fig:lir_sfr_t} should be placed side-by-side} This represents a
20\% reduction in $\lir$ vs.\ SFR scatter compared to $t_{\rm
SF}=10^8$ SF timescale. The values of scatter we give here are for the
$\lir$ to SED SFR ratio, but very similar answers are obtained if we
consider a scatter around the best linear fit.

\begin{figure}
\epsscale{1.2}
\plotone{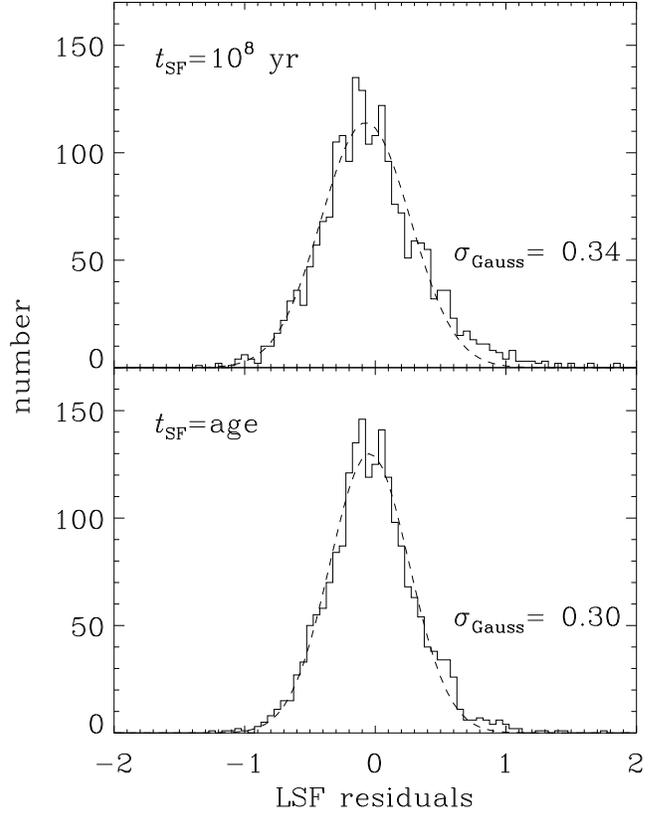}
\caption{Distribution of the residuals of the linear fit of infrared
luminosities and dust-corrected UV/optical star formation rates. Upper
panel shows residuals with respect to SFRs averaged over $10^8$ yr,
and the lower with respect to SFRs averaged over the population age. A
Gaussian is fit to each distribution (dashed curve), and its width
($\sigma_{\rm Gauss}$) is displayed. The Gaussian of the residuals
with respect to SFRs averaged over the population age is narrower. The
horizontal axis is in decades (dex).}
\label{fig:lir_sfr_gauss}
\end{figure}

One may get an impression that most of the reduction in scatter as we
go to longer SFR-averaging timescales is due to fewer outliers. To
check this, in Figure \ref{fig:lir_sfr_gauss} we fit Gaussian
functions to residuals around the linear fits, and find that the width
of the Gaussian (which is not dominated by outliers) decreases
similarly as the overall scatter, indicating that the decrease in
scatter is not due to the decrease in the number of outliers.

The most pressing concern with the above result is that the reduction
in scatter when comparing $\lir$ to SFR over progressively longer
timescales is an artifact of the SED fitting procedure, such that it
simply reflects the precision with which we are able measure SFRs (and
dust corrections) at different timescales (or alternatively,
wavelengths)? The average formal error in our SFR measurements is
between 0.14 and 0.18 dex for SFRs averaged over timescales $10^8$ yr
and longer. The small value of the error and its small variation for
different timescales implies that the SFR uncertainties are not
modulating the level of correlation with $\lir$, i.e., the change in
scatter is not driven by errors in the {\it measurement} of the
SFR. While the above is true on average, in \S \ref{sec:sed} we saw
that in some cases the error in SFR averaged over $10^8$ yr can get
relatively high. Removing all galaxies with error larger than 0.2 dex
leads to some reduction in scatter with respect to $\lir$, but is
still larger than the scatter between $\lir$ and age-averaged SFR. The
same is true if we limit the sample only to objects where the error in
SFR averaged over $10^8$ yr is smaller than the error of age-averaged
SFR.

To further test if we are able to reliably measure the SFR on $10^8$
yr timescale, we run simulations described in Appendix A. From those
we conclude that if the IR luminosity were indeed the reflection of
the current SF, then our UV/optical SFR averaged over $10^8$ yr {\it
would} measure it with a {\it smaller} scatter than an UV/optical SFR
averaged over any other longer timescale.

Finally, we notice that the average offset with respect to Kennicutt
conversion is larger when age-averaged SFRs are plotted instead of the
current ones in Figure \ref{fig:lir_sfr_t}. This should not be
surprising since the Kennicutt conversion was {\it calibrated}
assuming current ($\lesssim 10^8$ yr) SFR, and in general the SFR
averaged over longer time periods will be higher than the current one
(because most galaxies have declining star formation
histories). Additionally, the correlation is now steeper (the slope
from the bisector linear fit was 1.06 for $t_{\rm SF}=10^8$ and is now
1.20). Since in both cases one has the same $\lir$, the change in
slope has to be the result of a differential change in SFR between the
two averaging timescales for galaxies with low and with high IR
luminosities. As can be seen from Figure \ref{fig:lir_z}, galaxies
with $\lir< 10^{10} L_{\odot}$ are detected only at redshifts below
0.5. Galaxies that we see at this lower redshift will on average be
older than the galaxies observed at higher redshift, when the universe
was younger. For the same rate of SF decline, galaxies that had more
time to evolve ($\lir< 10^{10} L_{\odot}$ galaxies at $z<0.5$) will
show greater change between the age-averaged SFR and the current SFR
than the younger galaxies (more luminous). This moves $\lir< 10^{10}
L_{\odot}$ galaxies more to the right in Figure \ref{fig:lir_sfr_t}
than the more luminous ones, producing a steeper slope.

\subsection{$\lir$ and UV/optical SFR: dependence on galaxy color}

\begin{figure}
\epsscale{1.2} 
\plotone{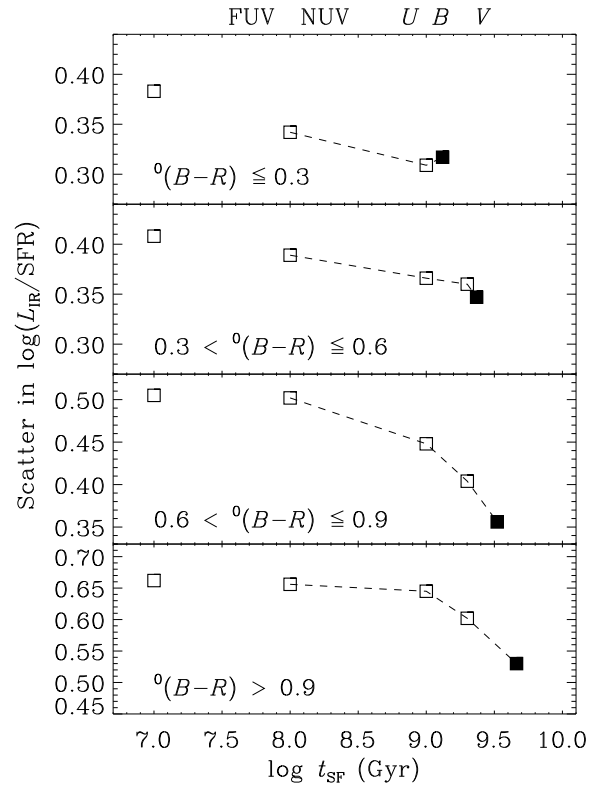}
\caption{Correlation between infrared luminosities and dust-corrected
UV/optical star formation as a function of SF averaging timescale, for
galaxies having different rest-frame colors. The top three panels
correspond to blue-sequence galaxies, and the bottom panel contains
green valley and red-sequence galaxies. The correlation generally gets
better (the scatter decreases) as the timescales increase. Filled
squares correspond to the age-averaged SFR and they are plotted at
$t_{\rm SF}$ corresponding to the average age of galaxies in a given
color bin. Unlike other timescales, the SFR averaged over $10^7$ yr is
poorly constrained in UV/optical SED fitting and would increase the
intrinsic scatter in $\lir$/SFR correlation. Each panel displays the
same relative dynamic range in vertical axis. Photometric bands
characteristic for a given timescale are given above the plot.}
\label{fig:lir_sfr_sct}
\end{figure}

The source of the IR luminosity will generally not be the same in
galaxies with different dominant stellar populations. Therefore, we
now explore the strength of the $\lir$ vs.\ SFR correlation, not only
as a function of timescale, but also for galaxies split into various
color bins. We now use rest-frame $B-R$ color because it somewhat
better discriminates the population age of star-forming galaxies than
${\rm NUV}-R$. In Figure \ref{fig:lir_sfr_sct} we plot the scatter of
the logarithm of $\lir$ to SFR ratio as a function of timescale over
which the SFR was averaged. Each panel displays the same relative
dynamic range in the vertical axis. The top three panels show
blue-sequence galaxies, while the bottom panel contains the green
valley and the red-sequence galaxies. Open squares represent the
scatter for corresponding fixed timescales, while the filled square is
the scatter in age-averaged SFR, plotted at the position of the
average age of galaxies in that color bin. Even for the bluest
galaxies ($^0(B-R)\leq0.3$, top panel), which have a large fraction of
recent star formation, the scatter is smallest at a timescale of
$10^9$ yr, rather than the $10^8$ yr UV timescale. In subsequent
redder color bins the best correlation with $\lir$ is always for
age-averaged SFR, i.e., on timescales of $\sim$ 2--3 Gyr. We tried to
identify a galaxy population for which the IR luminosity {\it would}
best match the short timescale of $10^8$ yr. We looked at the
$\lir$/SFR scatter in bins of galaxy stellar mass ($M_*$), specific
SFR (SFR/$M_*$) and the age of the most recent burst.  We find that IR
best correlates with $10^8$ yr timescale only for $\log M_* <8.5$
galaxies. These are blue compact dwarfs that we can detect only out to
$z\sim0.4$, so the result is based on a small number of objects.

\begin{figure*}
\epsscale{1.0} 
\plotone{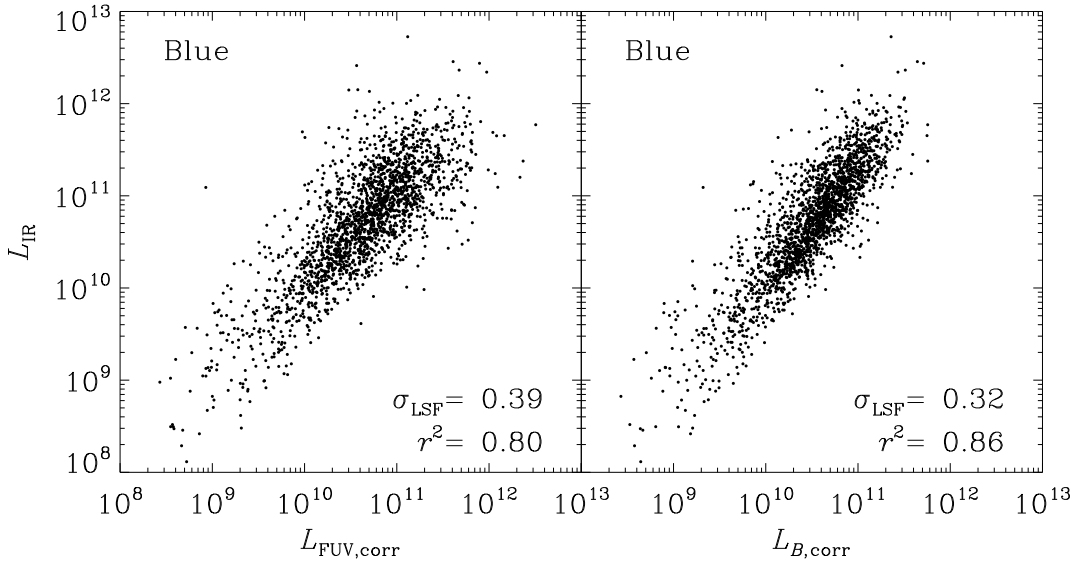}
\caption{Comparison of infrared luminosities and dust-corrected FUV
(left) and $B$-band luminosities (right) of blue-sequence
galaxies. Dust corrected FUV and $B$ luminosities have been derived
from the UV/optical SED fitting. Average correction is 2.2 mag in FUV
and 0.80 mag in $B$. The scatter around the least square linear fit is
given in the lower right corner, and is smaller for the $B$ band. Also
given is the correlation coefficient. These correlations are
equivalent to correlations between $\lir$ and the SFR averaged over
the short and long timescales presented in Figs.\
\ref{fig:lir_sfr_blue} and \ref{fig:lir_sfr_t}, but are more
fundamental in the sense that they tie $\lir$ with present-day source
of IR heating.}
\label{fig:lir_luvopt}
\end{figure*}

As explained previously, we begin our analysis using the concept of
SFRs averaged over various timescales. However, this was simply a
convenient way to probe {\it today's} stellar populations of different
ages. Given that the flux that is responsible for dust heating must be
produced at the present time, a quantity that will be more
fundamentally correlated to $\lir$ is some UV or optical
luminosity. For every SFR-averaging timescale $t_{\rm SF}$ there is a
characteristic (rest-frame) wavelength at which the population with
the age $t_{\rm SF}$ dominates. In Figure \ref{fig:lir_sfr_sct} we
show which of the bandpasses (FUV, NUV, $U$, $B$ and $V$) correspond
to various ages, extrapolated from \citet{oconnell}. The FUV will be
dominated by stars having ages $\lesssim 10^8$ yr, so the equivalent
to ``current'' SFR will be the {\it dust-corrected} FUV
luminosity. Similarly, the equivalent for SFR averaged over 1--3 Gyr
will be optical luminosity ($U$, $B$ or $V$), also corrected for
dust. The bandpass that corresponds to a timescale with the least
scatter in top panel of Figure \ref{fig:lir_sfr_sct} is between rest
frames $U$ and $B$, and around $V$ band for somewhat redder
blue-sequence galaxies (second and third panels). Now we again show IR
luminosities, but against dust-corrected UV or optical luminosities
instead of the SFRs. Figure \ref{fig:lir_luvopt} presents a comparison
of $\lir$ and dust-corrected FUV (left) and $B$ (right) luminosities,
again for blue-sequence galaxies.\footnote{Results for $B$-band
luminosity are very similar to those for $U$ or $V$, but we use $B$
since it is the most common band used for galaxy magnitudes.}  As in
the case of SFRs, the dust correction for FUV and $B$ band
luminosities come from our SED fitting, and it is on average 2.2 mag
in FUV and 0.8 mag in $B$. These figures are equivalent to those that
showed $\lir$ and SFR (Figs.\ \ref{fig:lir_sfr_blue} and
\ref{fig:lir_sfr_t}), and the arguments that applied for the
robustness of SFRs (Appendix A) apply here for FUV and $B$
luminosities. The scatter is smaller against $L_{B,{\rm corr}}$.
Formal scatter around the linear fit is 0.39 dex for FUV luminosity,
vs.\ 0.32 for $B$-band luminosity (0.36 and 0.30 dex when $3\sigma$
outliers are excluded). Pearson correlation coefficient provides
another way to measure the strength of a correlation. It is 0.80 for
FUV luminosity and 0.86 for $B$-band luminosity.  Note that in order
to have a meaningful comparison with IR luminosity, the UV or the
optical luminosity needs to be appropriately corrected for dust. In
absence of dust correction, the correlation coefficient between FUV
luminosity and $\lir$ drops to 0.59 and between $B$-band luminosity
and $\lir$ reduces slightly to 0.84 (since dust corrections in $B$ are
smaller).

The linear fit without $3\sigma$ outliers for $B$-band ($\lambda=4360$
\AA) is given by

\begin{equation}
\log \lir = 1.125(0.013)\, \log L_{B,{\rm corr}} -1.102(0.142), \qquad
\log L_{B,{\rm corr}}>8,
\label{eqn:lir_lb}
\end{equation}

\noindent where all luminosities are in $L_{\odot}$. The fit is
constructed for blue-sequence galaxies ($^0({\rm NUV}-R)<3.5$). The
values of parameters of the fit depend slightly ($\pm 0.05$ in slope)
on the exact color cut. The appropriate dust correction for $B$-band
can be obtained by scaling the attenuation in FUV using the mean
\citet{cf00} extinction law for age of $\sim1$ Gyr

\begin{equation}
A_B = 0.37 A_{\rm FUV},
\end{equation}

\noindent where $A_{\rm FUV}$ is preferably obtained from full SED
fitting, or alternatively using the UV slope relation given in
Equation \ref{eqn:uvslope} (but see discussion in Appendix B).

For some purposes, one may prefer a bisector linear fit \citep{isobe},
given by

\begin{equation}
\log \lir = 1.275(0.016)\, \log L_{B,{\rm corr}} -2.668(0.1673), \qquad
\log L_{B,{\rm corr}}>8,
\label{eqn:bis}
\end{equation}

\noindent which was constructed from all points in Figure
\ref{fig:lir_luvopt} (right).

The above correlation (Eqn.\ \ref{eqn:lir_lb} demonstrates that one
can essentially estimate, to within a factor of two, the total IR
luminosity from UV/optical photometry alone (i.e., the dust
corrections are also constrained only using UV/optical SED). While
this may not be the case for every type of galaxy at any redshift, it
appears true for normal and IR luminous star forming galaxies over the
$z<1.4$ redshift range studied here. Note that we obtain $\lir$ from
24 \mic\ flux, and that the correlation could perhaps be even tighter
with a better estimate of total $\lir$, one that employs longer
wavelength IR data and/or more accurate SED templates. This will be
addressed in future work.

\subsection{$\lir$ and UV/optical SFR: dependence on redshift} 
\label{ssec:lirz}

\begin{figure*}
\epsscale{1.0}
\plotone{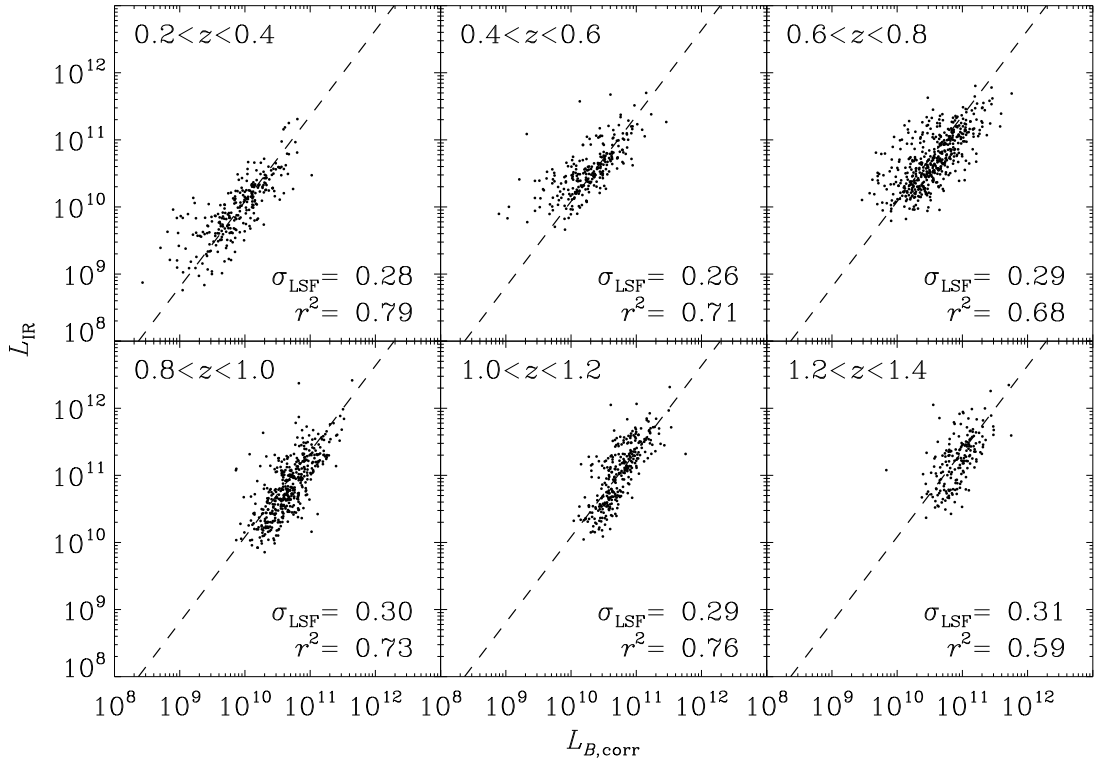}
\caption{Comparison of infrared luminosities and dust-corrected
$B$-band luminosity in different redshift bins (for blue-sequence
galaxies). Dust corrected $B$ luminosity has been derived from the
UV/optical SED fitting.  Numbers in lower right corners show the
dispersion around the least square linear fit (in dex) and the
correlation coefficient. The dashed line represent the bisector linear
fit for the entire sample, and is repeated from panel to panel as a
guiding line.}
\label{fig:lir_luvopt_z}
\end{figure*}

So far we have investigated the correlation between $\lir$ and SFR for
various samples irrespective of their redshift. Since the observed 24
\mic\ flux corresponds to 10--18 \mic\ rest-frame wavelength range
that contains both mid-IR continuum and strong PAH lines, one would
like to learn if there are any systematic differences in the $\lir$
vs.\ SFR correlation at different IR wavelengths, i.e., redshifts.
Still focusing on blue, star-forming galaxies, we find that the $\lir$
vs.\ SFR relation in different redshift bins follows the same trend as
found for the entire sample: $\lir$ correlates better with SFR
averaged over galaxy population age than over any shorter
timescale. Equivalently, and more fundamentally, $\lir$ correlates
better with dust-corrected optical luminosity than with dust-corrected
UV luminosity. This is shown in Figure \ref{fig:lir_luvopt_z} where we
present $\lir$ vs.\ dust-corrected $B$-band luminosity of blue
galaxies, split into six 0.2-wide redshift bins in the $0.2<z<1.4$
range. The upper number in the lower right corner of each panel is the
standard deviation around the least square linear fit, and the lower
number the correlation coefficient. The scatter is roughly the same in
all redshift bins, indicating that at this level of precision the
entire 10--18 \mic\ wavelength range corresponds equally well to the
UV/optical dust-corrected luminosity. The PAH features at 11.3 and
12.7 \mic\ would be sampled in $0.8<z<1.0$ and $1.0<z<1.2$ redshift
bins, respectively, and there we see a slight increase in scatter
compared to other redshift bins. In each panel we repeat the bisector
linear fit obtained for the full sample (Eqn.\ \ref{eqn:bis} as a
dashed line. From that one can see that the slope appears to get
steeper at higher redshifts. Rather than assuming that the intrinsic
linear relation changes at different redshifts, it is possible that
this is because the intrinsic relation is not linear. Then, since at
different redshifts one samples different ranges in luminosity,
segments of a curve will appear as linear relations with different
slopes. Also, it is plausible that the conversion from rest-frame
mid-IR flux to to $\lir$ has wavelength-dependent systematics.

One may wonder if the improvement in the correlation between
UV/optical luminosity and $\lir$ as we go to redder optical
luminosities (Fig.\ \ref{fig:lir_luvopt}) may in fact reflect a more
fundamental correlation of $\lir$ with stellar mass. On the one hand
this is not expected because dust heating should correlate with some
form of present flux, whatever from younger or older stars (or some
mix of two), and not on mass that includes stars that cannot
contribute significantly to the IR. On the other hand, we know that
for actively star forming galaxies the star formation rate and stellar
mass are tightly correlated (e.g., \citealt{boselli,b04}). When we
compare the $\lir$ of blue galaxies vs.\ their current stellar mass,
the scatter around the least square linear fit is 0.41 dex, and the
correlation coefficient is 0.76, which is weaker than what we found
when comparing $\lir$ to either FUV or $B$-band luminosity (corrected
for dust), with correlation coefficients 0.80 and 0.86,
respectively. Very similar results are obtained when we substitute the
current stellar mass with the estimate of total stellar mass formed
over the galaxy lifetime, i.e., the mass that includes
recycling. However, this is not the full story. There is evidence that
the SFR vs.\ mass relation evolves with redshift
\citep{papovich,noeske}), so for a given mass galaxies at different
redshifts will have different $\lir$. Indeed, if we split the sample
in 0.2 wide redshift bins (as in Fig.\ \ref{fig:lir_luvopt_z}), the
correlation between IR luminosity and the stellar mass improves, and
is comparable to that between IR luminosity and the dust-corrected
$B$-band luminosity in a given redshift bin ( Fig.\
\ref{fig:lir_luvopt_z}). Since the mass measurement in the SED fitting
is constrained by very similar information that constrain the optical
dust-corrected luminosity, this similarity between $M_*$ and $M_{B{\rm
,corr}}$ should not be surprising or considered or fundamental, but
instead reiterates the connection between the IR emission and the
stars other than the very young ones.

\section{Infrared luminosity and UV/optical SFR in green valley and 
red sequence galaxies} \label{sec:lir_red}

\begin{figure}
\epsscale{1.3}
\plotone{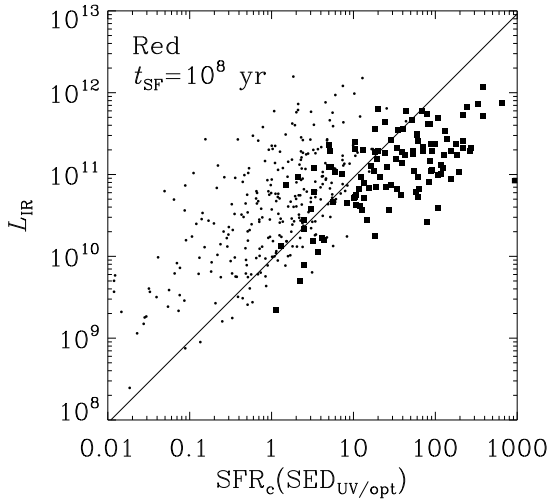}
\caption{Comparison of infrared luminosities and dust-corrected
UV/optical star formation rates for green valley and red sequence
galaxies.  Timescale for SED SFR is $10^8$ yr (subscript
'c'). Galaxies classified as dusty starbursts based on Figure
\ref{fig:ssfr_nuvr} are shown as filled squares. Other red galaxies
mostly lie above the \citet{kennicutt} conversion (solid line)
indicating that the current SF is not the primary source of IR
luminosity.}
\label{fig:lir_sfr_red}
\end{figure}

The analysis presented so far has focused on blue sequence galaxies,
for which it was reasonable to assume that IR emission would be
strongly related to active star formation. The picture becomes more
complex as one moves away from the blue sequence into the green valley
and the red sequence. While some galaxies in the green valley will
simply be reddened actively star-forming galaxies, others will have
such colors because they have little ongoing star formation
(corresponding to galaxies above and below the dashed line in Figure
\ref{fig:ssfr_nuvr}). We should expect older stellar populations to
contribute more to the IR emission in the latter group. This will be
even more the case for red sequence galaxies, which have little or no
current star formation. In Figure \ref{fig:ssfr_nuvr} we saw that
there exist 24 \mic-detected galaxies well into the red sequence (as
red as any galaxy in our optical sample). Their low specific SFRs
indicate that these galaxies are intrinsically very red and not just
dust reddened. In Figure \ref{fig:lir_sfr_red} we compare IR
luminosity and the SED-derived current SFR ($t_{\rm SF}=10^8$ yr) for
red galaxies (which includes the green valley and the red
sequence). We distinguish between dusty starbursts (galaxies above the
dashed line in Figure \ref{fig:ssfr_nuvr}, plotted as squares) and
regular red galaxies (dots). The two groups occupy distinct
locations. Dusty starbursts have high UV/optical SFRs: above 10
$M_{\odot}{\rm yr}^{-1}$, and in some cases approaching 1000
$M_{\odot}{\rm yr}^{-1}$. They lie close to the 1:1 \citet{kennicutt}
conversion between $\lir$ and SFR. This is expected if $\lir$ in dusty
starbursts is due to SF. Actually, galaxies with the most intense SF
have somewhat lower $\lir$ than the expected, ULIRG levels. For such
extreme cases it is possible that the SED fitting overpredicts the
dust correction (but we cannot exclude that IR luminosities are
perhaps underestimated). Non-dusty red galaxies (dots) have lower SFRs
and lie above the $\lir$--SFR conversion. This means that $\lir$ is
not powered by the current SF. At each UV/optical SFR there is a wide
range of IR luminosities. This again speaks of a disconnect between
$\lir$ and SFR.

\begin{figure}
\epsscale{1.2}
\plotone{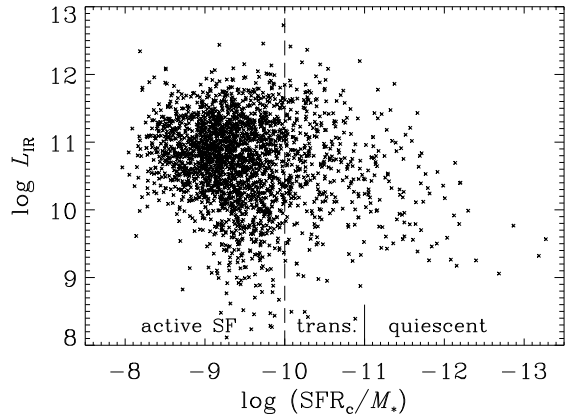}
\caption{IR luminosity of galaxies with different specific
SFRs. Actively star forming galaxies are to the left of the dashed
line, while those to the right are transitional or quiescent and they
correspond to green valley and red-sequence galaxies. LIRG-like
luminosities ($\log \lir>11$) can be found even among some very
quiescent galaxies.}
\label{fig:lir_ssfr}
\end{figure}

Non-dusty red galaxies are the main subject of the analysis in this
section. Can we explain the presence of 24 \mic\ emission and the
derived $\lir$ luminosities in these galaxies only with stellar
emission, which by necessity (since there is little current SF) will
mostly come from intermediate and old stellar populations? Do we see
evidence that some other dust heating mechanism, such as an AGN, may
be present in these galaxies? In Figure \ref{fig:lir_ssfr} the IR
luminosity for galaxies with different (current) specific SFRs is
shown. Since, unlike color, the specific SFR is corrected for dust,
this plot enables us to place red dusty starbursts together with blue
actively SF galaxies (left, $\log ({\rm SFR}/M_*)>-10$), and separate
more quiescent, red galaxies (right, $\log ({\rm SFR}/M_)*<-10$) for
which we investigate the source of IR emission. We see that galaxies
with LIRG-like luminosities are present well beyond the region of
actively star-forming galaxies, with some having specific SFRs as low
as $\log ({\rm SFR}/M_*)=-11.5$, which corresponds to rest-frame color
of ${\rm NUV}-R=5$, the color of the bluest nearby elliptical galaxies
\citep{donas}. One can be concerned that the use of IR templates based
on actively SF galaxies to derive $\lir$ for these more quiescent
objects may not be appropriate. This is entirely possible. However, as
in the previous analysis, we will assume that this (commonly used)
procedure is correct and then draw consequences. At each specific SFR
there is a wide range of $\lir$, especially for active galaxies. This
is mostly the consequence of a wide range of masses probed at each
specific SFR.

\begin{figure}
\epsscale{1.2}
\plotone{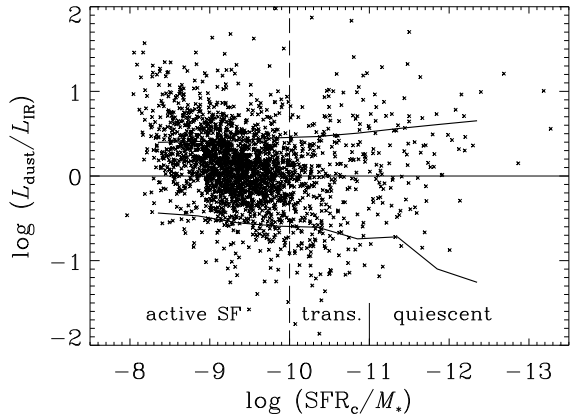}
\caption{Ratio of dust luminosity to the observed IR luminosity
against the current specific SFR. $L_{\rm dust}$ is stellar energy
absorbed by the dust and is derived from the UV/optical SED
fitting. Most galaxies, both actively star forming (left of the dashed
line) and quiescent (together with transitional; right of the dashed
line) have ratios around unity (horizontal line), indicating that the
dust absorption of stellar light can on average account for the
observed $\lir$. The two thick line represent $\pm 1\, \sigma$ range
of error in the ratio.}
\label{fig:ldustlir_ssfr}
\end{figure}

In order to establish if the IR luminosities that we see in red
galaxies can be produced by stars alone (of any age) we perform the
following exercise. The UV/optical SED fitting allows us to estimate
the total amount of stellar luminosity {\it absorbed} by the dust
\citep{cortese,dacunha}. According to the dust model of \citet{cf00},
this energy ({\it dust luminosity}), will come from birth clouds
surrounding young stars ($<10$ Myr old) and from the ISM heated by
stars of intermediate and older age. In the case when there is no
non-stellar source of IR emission, the $\lir$ should match, or at
least not significantly exceed the $L_{\rm dust}$ estimate. Estimating
the amount of dust extinction in quiescent galaxies from the
UV/optical SED fitting will be more uncertain than in actively
star-forming galaxies, as suggested in Figure
\ref{fig:sfrerr}. Nonetheless, we expect that the dust luminosity
derived from the SED fitting should on average be correct, which
therefore allows us to check the energy budget. In Figure
\ref{fig:ldustlir_ssfr} we present the ratio of the dust luminosity
($L_{\rm dust}$) derived from the UV/optical SED fitting to the
observed IR luminosity against the current specific SFR. Objects to
the right of $\log ({\rm SFR}/M_*)=-10$ line are red, quiescent (or
transitional) galaxies. Ratio of $L_{\rm dust}$ to $\lir$ of unity
means that the energy that is estimates to be absorbed in the
UV/optical part of the spectrum equals the energy re-emitted in the
IR. To see what range of values are consistent with the ratio of 1, we
calculate average 68\% confidence range of the ratio (two thick lines)
from PDF errors of $L_{\rm dust}$, with an {\it ad hoc} 0.3 dex error
for $\lir$ added in quadrature. Most of the actively SF galaxies fall
in the region consistent with the ratio of unity except those with
high specific SFRs. As already mentioned, for these objects SED SFR
(and therefore $L_{\rm dust}$) may be overestimated, or their $\lir$
underestimated. Red quiescent galaxies have a larger scatter of
$L_{\rm dust}/\lir$ ratios, which is not surprising given the higher
uncertainties in estimating $L_{\rm dust}$ from the SED fitting for
these galaxies (the thick lines). Again, most galaxies lie within the
$\pm1\,\sigma$ range around unity. From this we conclude that the dust
heated by stellar populations is roughly sufficient to account for the
observed IR luminosity even for relatively quiescent galaxies with
LIRG-like IR luminosities. Consequently, we conclude that there cannot
be a large population of (presumably obscured) AGNs which would
significantly raise $\lir$ and skew the ratio below unity (we will
next see that AGN may be affecting $\lir$, but only at a moderate
level). Given the low levels of current SF in transitional and
quiescent galaxies, one must conclude that intermediate and older
stellar populations produce the bulk of the IR emission (see also
Figure \ref{fig:ldustyo}).

\begin{figure}
\epsscale{1.2}
\plotone{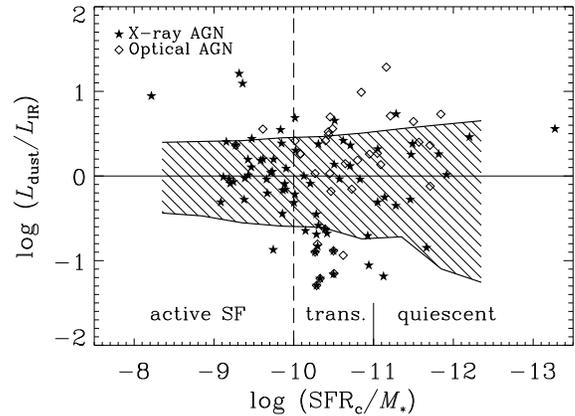}
\caption{Ratio of dust luminosity and the observed IR luminosity
against the current specific SFR for galaxies identified as AGNs. AGNs
are identified from X-ray detections (star symbols) and from optical
emission lines (open diamonds). Relatively large number of AGNs,
especially optically identified, have transitional or quiescent SF
histories (right of the dashed line). Most AGNs have $L_{\rm dust}$ to
$\lir$ ratio consistent with unity (the shaded region between two
thick lines represent $\pm 1\, \sigma$ range of error in the ratio,
based on the full sample (Fig.\ \ref{fig:ldustlir_ssfr})). This
suggests that AGN heating of the dust is on average not very
significant. Exceptions may be AGNs lying below the lower thick line
with $-11<\log ({\rm SFR}/M_*)<-10$.}
\label{fig:ldustlir_ssfr_agn}
\end{figure}

Regardless of the arguments laid out above, one would still like to
test directly whether the presence of AGNs has a significant effect on
the mid-IR emission in our sample, especially among the more quiescent
galaxies. Obtaining a full census of AGNs in our sample is not
straightforward. First, our UV/optical SFRs can be derived only for
galaxies where an AGN has no effect on the UV continuum, which is why
we have already excluded several tens of broad-line AGNs (type 1 AGN),
as identified from the spectra. To identify narrow-line AGNs (type 2
AGN) using the BPT emission line classification (BPT, \citealt{bpt})
requires spectra that cover rest frame range from 4800--6600 \AA. For
our spectra this is possible only in a very small redshift range
($0.33<z<0.38$).  However, coupled with information on stellar mass,
some fraction of galaxies lying in the AGN parts of the BPT diagnostic
diagram can be distinguished even in single-axis projections of the
diagram, i.e., using one line ratio. Using similar criteria as
\citet{weiner}, we select AGN candidates at $z<0.38$ by requiring the
flux ratio $\log{\rm([NII]/H\alpha})>-0.2$ and stellar mass $\log
M_*>9.5$, and at $0.34<z<0.82$ by selecting
$\log{\rm([OIII]/H\beta})>0.7$ together with $\log M_*>10.2$. These
criteria select a total of 35 type 2 AGN candidates detected at 24
\mic, which we call ``optical AGN''. Additionally, we use a catalog of
{\it Chandra} sources \citep{laird} to identify 74 X-ray AGN
candidates detected at 24 \mic. The majority of X-ray sources in the
EGS are believed to be AGNs or have an AGN component \citep{laird}. We
plot the ratio of $L_{\rm dust}$ to $\lir$ against the current
specific SFR in Figure \ref{fig:ldustlir_ssfr_agn} coding points by
AGN type. First, we notice that optical AGNs (open diamonds) are
almost exclusively transitional or quiescent objects. This agrees with
the results of local studies where there appears to be a relation
between optical AGN and SF quenching
\citep{kauffmann,s07,graves,schawinski}. X-ray AGNs are additionally
present among the galaxies with higher specific SFRs, but only up to
$\log ({\rm SFR}/M_*)=-9$, which again may be related to their role in
SF quenching. Most AGNs have $L_{\rm dust}/\lir$ consistent with unity
(the $\pm1\,\sigma$ range, shaded, is repeated from Figure
\ref{fig:ldustlir_ssfr}). If AGN contributes significantly to $\lir$
this would be reflected in $L_{\rm dust}/\lir$ below that of non-AGN
galaxies. While this is generally not the case, there is a group of
AGN at $-11<\log ({\rm SFR}/M_*)<-10$ where in {\it individual cases}
the AGN contribution to $\lir$ may be around 90\%.

\begin{figure}
\epsscale{1.2}
\plotone{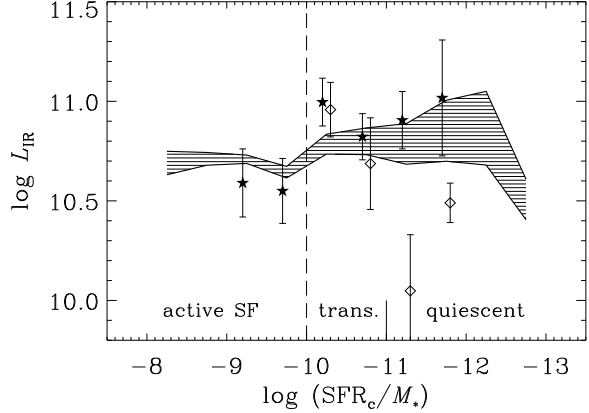}
\caption{Average $\lir$ of AGNs and non-AGNs in bins of specific
SFR. Plotted are: 1) non-AGNs (shaded region between two thick lines),
2) X-ray AGNs (star symbols) and 3) optical AGNs (open
diamonds). Error bars give the error of the mean in each bin. The
error range for non-AGN is given by the shaded region. X-ray AGNs are
somewhat more luminous than non-AGNs around $\log ({\rm
SFR}/M_*)=-10.3$. Two sets of error bars are slightly offset between
each other in horizontal direction for clarity.}
\label{fig:lir_ssfr_avecomp}
\end{figure}

Next we try to estimate the fraction of $\lir$ that is {\it on
average} attributable to AGNs. Figure \ref{fig:lir_ssfr_avecomp}
displays average IR luminosities in bins of current specific SFR for
three classes of galaxies: 1) non-AGNs (shaded region between two
thick lines), 2) X-ray AGNs (star symbols) and 3) optical AGNs (open
diamonds). Error bars give the error of the mean in each bin. On the
actively SF side we have only X-ray AGNs, and their average $\lir$ is
consistent with those of non-AGNs. On the quiescent side X-ray AGN
have $\lir$ up to 0.2 dex higher than non-AGNs, although it is only at
$\log ({\rm SFR}/M_*)=-10.3$ that the excess is somewhat
significant. Optical AGNs are similar to X-ray AGNs at transitional
specific SFRs, and then significantly lower than non-AGNs for low
specific SFRs, most probably because these are low-power AGNs such as
LINERs.

The above procedure has a drawback that if AGN selection is biased
with respect to $\lir$, their average $\lir$ will be off. Thus we
append it with the following test. For each group of AGNs (optical and
X-ray) we select a control group of non-AGNs with similar
properties. This needs to be done for {\it all} AGN (47 optical and 86
X-ray) regardless of whether they have been detected at 24 \mic. For
each optical AGN we select an object from the same redshift range
($z<0.38$ or $0.34<z<0.82$) such that the emission lines do not
indicate an AGN, and with a matching stellar mass and specific
SFR. The matching object is defined as the one that minimizes the
``distance'' $D$ in the stellar mass--specific SFR space:

\begin{equation}
D^2 = (\Delta \log M_*)^2 + c^2\Delta(\log ({\rm SFR}/M_*))^2,
\label{eqn:distance}
\end{equation}

\noindent where $c$ is a ``scaling'' ratio between $\log M_*$ and
$\log ({\rm SFR}/M_*)$, which we nominally take to be 3 based on the
range of these quantities in our sample of AGN. Similarity in stellar
mass and current specific SFR will ensure similarity in many other
non-AGN characteristics as well \citep{schiminovich}. For X-ray AGNs
we select matching non-AGNs such that they are not detected in X-ray,
have a redshift within 0.2, and minimize Equation
\ref{eqn:distance}. For both samples the same non-AGN match is allowed
to appear more than once.

\begin{figure}
\epsscale{1.2} \plotone{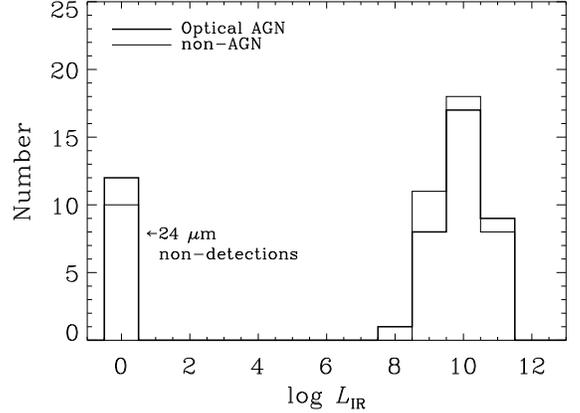}
\caption{Infrared luminosity distribution of type 2 AGNs selected
using a single-line BPT diagram (optical AGN, thick histogram), and of
the control sample of non-AGNs (thin histogram). Non-detections at 24
\mic\ are plotted at $\log \lir=0$ in both case. Each group contains
47 objects. Non-AGNs were selected to have similar masses and specific
SFRs as AGNs. There is no significant difference in IR luminosities
between the two groups, or in the number of 24 \mic\ non-detections
(12 for AGN, 10 for non-AGN.}
\label{fig:bptagn_lir}
\end{figure}

\begin{figure}
\epsscale{1.2} \plotone{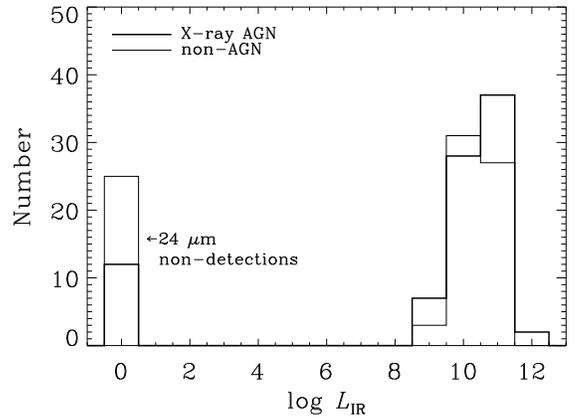}
\caption{Infrared luminosity distribution of X-ray detected AGNs
(thick histogram), and of the control sample of non-AGNs (thin
histogram). Non-detections at 24 \mic\ are plotted at $\log \lir=0$ in
both case. Each group contains 86 objects. Non-AGNs were selected to
have similar masses and current specific SFRs as AGNs. Non-AGN are more
frequently (24 vs.\ 12 galaxies) not detected at 24 \mic, making them
on average somewhat less luminous in IR.}
\label{fig:xrayagn_lir}
\end{figure}

In Figure \ref{fig:bptagn_lir} we compare the distribution of IR
luminosities for optical AGNs (thick histogram) vs.\ non-AGNs (thin
histogram). Objects not detected at 24 \mic\ are plotted with
$\log\lir=0$. The two distributions are quite similar, including the
similar number of 24 \mic\ non-detections. The average $\lir$ of the
AGN is 0.16 dex higher than that of the non-AGNs (for the part of the
sample where AGN and non-AGN are both detected at 24 \mic). However,
the average stellar mass of the AGN sample is also slightly higher
(0.24 dex), making the difference in $\lir$ less significant. Results
are similar when we choose $c=2$ or 4 in Equation
\ref{eqn:distance}. From this we conclude that optical AGNs are drawn
from the same underlying IR population as non-AGNs. A similar
comparison is shown for X-ray AGNs in Figure \ref{fig:xrayagn_lir}. If
both an AGN and a matching non-AGN are detected in 24 \mic, their
$\lir$ (and stellar mass) are on average quite similar, as was the
case with optical AGNs. However, while 12 X-ray AGN are not detected
in 24 \mic, this number jumps to 24 for the non-AGN control group. If
we assume that each non-detection has $\lir$ corresponding to the
detection limit at the given redshift, we get that the average $\lir$
of X-ray AGNs is 0.23 dex higher than of non-AGNs (both AGNs and
non-AGNs have the same stellar masses). This result is robust if we
choose $c=2$ or 4 in Equation \ref{eqn:distance}.

While our sample of AGN is not large enough to draw firm conclusions,
it appears that AGN are not a significant contributor to mid-IR
luminosities in the general case, i.e., in samples that have an
optical selection, such as ours. Where their presence could be
detected, especially among the transitional and quiescent galaxies,
they still contribute at most 50\% of $\lir$. The contribution of AGN
$\lir$ from these galaxies to the global SFR density is beyond the
scope of this work, but is most likely not very high.

\section{Discussion} \label{sec:disc}

\subsection{Dust heating in actively star-forming galaxies}

\begin{figure}
\epsscale{1.2}
\plotone{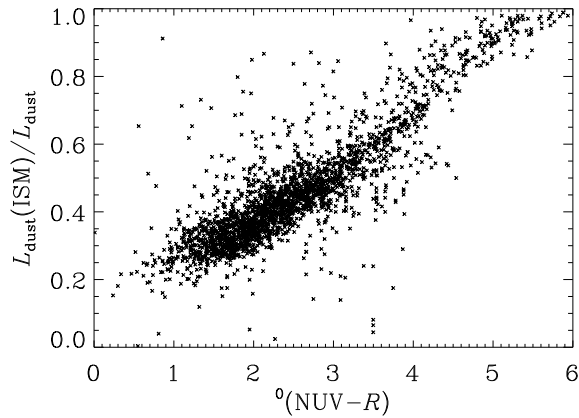}
\caption{Fraction of the dust luminosity due to ambient ISM as a
function of rest-frame galaxy color. Total dust luminosity includes
includes energy absorbed in birth clouds ($<10$ Myr) plus the ISM. The
ratio is constrained by dust prescription of \citet{cf00}. Even
blue-sequence galaxies can have a high fraction of their dust
luminosity absorbed away from birth clouds.}
\label{fig:ldustyo}
\end{figure}

Analysis presented in \S \ref{sec:lir_sfr} indicates that the IR
luminosity extrapolated from mid-IR flux is better correlated with the
optical light of intermediate age populations than with the UV light
of young stars. This result can be interpreted as the larger
contribution of intermediate-age stars than of young stars in the
mid-IR dust heating. Such interpretation is at odds with recent
studies on nearby galaxies that find very good correlation between
nebular emission line Pa$\alpha$, that comes from massive young stars
($<10$ Myr old), and the rest-frame 24\mic\ luminosity
\citep{calzetti07,alonso,rieke}. Note, however, that we explore a
different part of the mid-IR wavelength range (10--18 \mic) which is
more affected by PAH emission, and could therefore be more strongly
correlated with the cold, diffuse dust from older stellar
populations. What fraction of total $\lir$ (irrespective of IR
wavelength range) is expected to come from young stars in this
particular sample? The dust model that we use in our SED fitting
\citep{cf00} allows us to estimate the relative contribution of the
stellar energy absorbed by the stellar birth clouds (emitted by young
stars) and the ambient ISM (emitted by intermediate age and old
stars). This is achieved simply by considering the UV/optical
luminosity that is absorbed in these two components. In Figure
\ref{fig:ldustyo} we plot the fraction of dust luminosity contributed
by the ISM, i.e., away from the sites of current star formation. Not
surprisingly, this fraction correlates well with NUV$-R$ color, which
to first order gives the ratio of the recent to past star
formation. For blue, star-forming galaxies ($^0({\rm NUV}-R)<3.5$) the
fraction of dust heating, and therefore the $\lir$ due to ISM can be
as high as 60\% and is typically 40\%. While still not dominant, the
ratio of dust heating due to ISM can thus be quite significant. Note
that the ratio presented here does not constitute a measurement, but
is set by the dust model we use here. However, this dust model is
physically motivated and can therefore serve as a guide. In reality,
the contribution of the ISM may be somewhat different. Generally,
there are many uncertainties with respect to the evolution of the
intermediate age population and their dust production to leave room
for their greater contribution to dust luminosity, perhaps even at
mid-IR wavelengths.

\subsection{Dust heating in quiescent galaxies}

The presence of high IR luminosities in galaxies that appear quiescent
(not actively star-forming) based on UV/optical SEDs seems
puzzling. Since for a given current SFR we have such a wide range of
IR luminosities (including a large number of optically luminous, yet
24 \mic-undetected galaxies [Fig.\ \ref{fig:nuvr_hist}]), the
disconnect between the star formation and the IR properties appears
strong.  We have shown that AGNs, while contributing, do not dominate
in the mid-IR, and that the absorbed stellar emission (from
intermediate age and older stars) can on average reproduce high IR
luminosities. Yet, this requires attenuations that are significantly
higher than what we see in nearby galaxies with similarly low levels
of specific star formation. Locally, such galaxies are morphologically
early-type galaxies, with low to moderate amounts of dust and with
infrared luminosities that do not exceed $10^{10} L_{\odot}$
\citep{gdj}. Cursory examination of optical {\it HST} ACS images of
galaxies in our sample with $\log \lir>10$ and $^0({\rm NUV}-R)>5$
(corresponding to UV/optical colors of nearby ellipticals,
\citealt{donas}) indicates that 2/3 of them indeed look like
early-type galaxies, and the rest are either edge-on disks, or show
some structure. These results indicate that a fraction of early-type
galaxies at higher redshifts have significantly higher dust contents,
leading to higher IR luminosities. The presence of large amounts of
dust even in nearby ellipticals is an open question \citep{temi}, and
is outside of the scope of this work.

Another explanation for apparent high IR luminosities is that because
we use IR templates based on star-forming galaxies to estimate the
total $\lir$ of more quiescent galaxies, that this leads to
significant overestimates. This explanation is not intuitive since one
expects quiescent galaxies to have colder dust and therefore the
mid-IR flux point used in conjunction with star-forming templates to
underestimate the total $\lir$. But this explanation could be valid if
quiescent galaxies contained significant contribution of stellar
components that peak in the mid-IR (and are not included in IR
templates), such as the dust around the AGB stars \citep{bressan}.

\subsection{Monochromatic IR luminosity and SFR} \label{ssec:mono}

Recently there have been efforts to explore the use of {\it
monochromatic} mid-IR luminosity as a tracer of star formation, either
as a substitute for the total IR luminosity or as a measure that is
intrinsically better correlated with SFR
\citep{calzetti07,rieke}. Therefore, for our observations at 24 \mic,
we construct a luminosity estimate at 12 \mic\ ($L_{12}$), which
corresponds to rest-frame of our observations at $z=1$. To get
$L_{12}$ for galaxies at other redshifts, we again rely on IR
templates \citep{dh02}, but now to obtain only a relatively small
K-correction, instead of a full bolometric correction. Comparing
$L_{12}$ to FUV and $B$-band dust-corrected luminosities we find that
the linear fits have scatters of 0.34 and 0.28 dex, respectively,
i.e., they are some $\sim15\%$ smaller than in relations with
$\lir$. However, this comparison can be misleading since the {\it
range} of $L_{12}$ values is different (smaller) than of $\lir$. If
instead we compare Pearson correlation coefficients, we find that they
are basically the same for $L_{12}$ and for $\lir$. While this does
not mean that the total luminosity is intrinsically better correlated
with the UV/optical SFR than the 12 \mic\ luminosity, it does at least
argue that at our level of precision the correction to total
luminosity neither does introduce significant additional uncertainty,
nor does it offer any measurable benefits.

\subsection{IR excess and Compton-thick AGN}

Recently, \citet[D07]{daddi} have studied a population of $z\sim2$
galaxies in GOODS fields that exhibits a mid-IR excess (IRX) around 8
\mic and ascribed this excess to heating from Compton-thick AGNs. In
\S \ref{sec:lir_red} we argued against the need for non-stellar
sources of IR heating, yet one would like to explore if there is a
population of similar mid-IR-excess sources in our sample (at 10--18
\mic). D07 define mid-IR-excess objects as those with the ratio of the
combined IR and UV SFR (the latter not corrected for dust) to
dust-corrected UV SFR exceeding 3.16 ($10^{0.5}$):

\begin{equation}
{\rm IRX(D07)} = {\rm SFR_{IR+UV}/SFR_{UV,corr}} >
3.16. \label{eqn:irx}
\end{equation}

\noindent D07 obtain IR SFR by extrapolating 24 \mic\ flux (which
corresponds to $\sim 8$ \mic\ rest-frame flux) to total IR luminosity
and then using the \citet{kennicutt} SFR conversion. They get
dust-corrected UV SFR by applying a correction based on the fixed
relation between the UV slope and UV attenuation. We will construct
the IRX measure in exactly the same way, except that our $\lir$ is
extrapolated from 10--18 \mic\ rest-frame flux. Using models based on
\citet{bc03} we find the following relation (``K-correction'') between
the observed $B-z$ color at $z=2$ and the rest-frame UV color:

\begin{equation}
(B-z)_{z=2}=1.8\,^0({\rm FUV}-{\rm NUV}).
\end{equation}

\begin{figure}
\epsscale{1.2}
\plotone{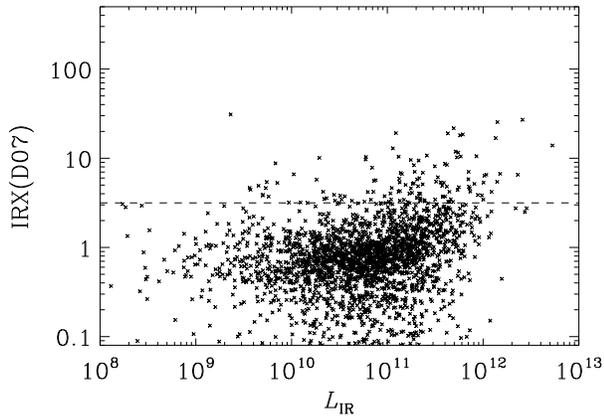}
\caption{Mid-IR-excess (IRX) calculated according to \citet{daddi} as
a function of IR luminosity. IRX is defined as the ratio of SFR summed
from $\lir$ plus uncorrected FUV SFR to dust-corrected FUV SFR (see
text for details). This plot should be compared to Fig.\ 2 of
\citet{daddi}, which shows many $z\sim 2$ IR-excess objects
(IRX$>3.16$, above the dashed line) with ULIRG luminosities, while
there are few such objects in this sample. Nevertheless, we find some
objects with moderate IR excess. The large majority of AGNs (88\%) are
not IR-excess objects, although AGNs are somewhat more frequent among
IR-excess objects (10\%) than among those that are not (4\%).}
\label{fig:irx}
\end{figure}

\noindent Therefore their relation between the dust reddening and the
observed $B$ and $z$ magnitudes at $z=2$ (\citealt{daddi1}, Eqn.\ 8)
corresponds to FUV attenuation of $A_{\rm FUV}= 4.5\,({\rm FUV}-{\rm
NUV})$, which is similar in slope to that proposed by \citet{meurer}
for star-bursting galaxies. In Figure \ref{fig:irx} we plot IRX
against $\lir$ for our galaxies. This figure should be qualitatively
compared to D07 Fig.\ 2 except that the horizontal axis in D07 shows 8
\mic\ rest-frame luminosity ($L$(8\,\mic)) while our figure shows
total $\lir$. The $\lir$ range of our sample of $9<\log \lir<12$
roughly translates into the range $8.5<\log L(8\mu{\rm m})<11$ range
\citep{daddi1}. The dashed line designates the IR excess criterion of
D07. The vast majority of galaxies in our sample show no IR excess
(IRX(D07)$\approx 1$). A small number of galaxies in our sample are
found above the limit. Unlike in D07 where IR-excess objects comprise
nearly all high-luminosity objects, here there is almost no range in
IR luminosity where excess objects dominate. Also, the maximum levels
of IR excess in our sample are around 30, while in D07 they are up to
10 times higher. This is probably related to the very different nature
of the two samples. Majority of the extreme IR-excess objects of D07
have $\log L(8\mu{\rm m})>11$, which corresponds to ULIRG
luminosities. Such objects are all but absent in our sample. Indeed, a
few objects that have $\lir>10^{12} L_{\odot}$ in our sample have a
moderate excess. Trying to determine if the IR-excess objects in our
sample harbor Compton-thick AGN is beyond the scope of this
paper. Some recent work suggests alternative explanations for mid-IR
excess at $z\sim2$ involving PAH emission at 8 \mic
\citep{murphy,huang}. The AGN that we identify in Section
\ref{sec:lir_red} are mostly not IR-excess sources (only 12\% of
optical or X-ray AGN have IRX$>3.16$), although they do represent
somewhat higher fraction among IR-excess sources (10\%) than among
those which are not (4\%). This is consistent with conclusions of
Section \ref{sec:lir_red} that on average AGNs contribute moderately
to the mid-IR flux, unlike the possibly dominant AGN contribution
among the IR luminous mid-IR excess sources of D07. Also, note that
Eqn.\ \ref{eqn:irx}, by applying the Kennicutt conversion between IR
luminosity and SFR, {\it assumes} that IR luminosity is dominated by
young populations. This is certainly not true for more quiescent
galaxies present in our sample.

\section{Conclusions}

In this paper we present (1) SFRs based on the Bayesian SED modeling
of the UV and optical stellar continuum emission, obtained by applying
the \citet{cf00} dust attenuation model to a suite of \citet{bc03}
stellar population synthesis models, and (2) the total infrared
luminosities extrapolated from 24 \mic\ observations using the SED
templates of \citet{dh02}, calibrated with local luminosity--color
relations. We cover the redshift range up to $z= 1.4$, and study
galaxies with spectroscopic redshifts from the AEGIS survey (Fig.\
\ref{fig:area}). Our sample of 24 \mic-detected $R$ band-selected
objects contains normal star forming galaxies and LIRGs, as well as
quiescent galaxies, and is not biased against IR-luminous populations
(Figs.\ \ref{fig:ssfr_nuvr} and \ref{fig:lir_z}). We compare IR
luminosity with UV/optical SFRs averaged over various timescales, thus
probing the present-day luminosities of stellar populations ranging in
age from 0.1 to 3 Gyr. From this analysis we conclude the following:

1. \enspace When comparing UV/optical SFRs to IR luminosities, we
   confirm that one needs to treat actively star forming galaxies
   separately from more quiescent ones. This caveat is well known in
   the study of nearby galaxies (e.g., \citealt{kennicutt}) but is
   sometimes neglected at higher redshifts. Points 2-6 below pertain
   to actively SF galaxies which we select using a cut on rest-frame
   ${\rm NUV}-R$ color (Figs.\ \ref{fig:nuvr_hist} and
   \ref{fig:ssfr_nuvr} .

2. \enspace UV/optical SFRs averaged over relatively short timescales
   ($10^8$ yr), and thus representing current SFRs, compare well
   (average difference 0.03 dex) with $\lir$ converted into SFR using
   the \citet{kennicutt} conversion. However, the scatter between such
   SFRs and $\lir$ is relatively high (0.42 dex) (Fig.\
   \ref{fig:lir_sfr_all}).

3. \enspace The scatter between UV/optical SFRs and $\lir$ reduces as
   one considers SFRs averaged over longer periods of time, and is
   best for timescales between 1 and 3 Gyr, depending on the color
   (i.e., dominant population age) of a galaxy (Figs.\
   \ref{fig:lir_sfr_blue}--\ref{fig:lir_sfr_sct}. Equivalently, but
   more fundamentally, this means that the correlation is better
   between $\lir$ and $B$-band dust-corrected luminosity (0.32 dex)
   than against FUV dust-corrected luminosity (0.39 dex) (Fig.\
   \ref{fig:lir_luvopt}). This argues for a significant role of
   intermediate-age stellar populations in mid-IR heating.

4. \enspace Better correlation of $\lir$ with optical luminosity than
   with FUV luminosity holds in redshift bins throughout $0.2<z<1.4$,
   corresponding to 10--18 \mic\ rest-frame wavelengths (Fig.\
   \ref{fig:lir_luvopt_z}).

5. \enspace For our sample, which mostly consists of LIRGs and normal
   star forming galaxies, we find that galaxies are on average not
   optically thick, i.e., their IR luminosity (extrapolated from the
   24 \mic\ flux) can be estimated from UV and optical photometry to
   within a factor of two (Fig.\ \ref{fig:lir_luvopt}) .

6. \enspace Many green valley galaxies are simply dust-obscured
   actively star-forming galaxies. However, there exist 24
   \mic-detected galaxies, some with LIRG-like luminosities, which
   have little current star formation (low specific SFR), i.e, they
   belong to green valley and even the red sequence because of their
   star formation history, not just dust reddening (Figs.\
   \ref{fig:lir_sfr_red} and \ref{fig:lir_ssfr}). 

7. \enspace On average, modeled amounts of dust absorption of stellar
   light are sufficient to produce the observed levels of IR
   luminosity, both for blue and for red-sequence galaxies (Fig.\
   \ref{fig:ldustlir_ssfr}). For red, quiescent galaxies this must
   include a large contribution of intermediate and old stellar
   populations and higher dust attenuations than in nearby early-type
   galaxies.  \ref{fig:ldustlir_ssfr}).

8. \enspace Identified AGNs on average do not contribute significantly
   to mid-IR luminosity at these redshifts. We see no evidence for a
   contribution by optical (type 2) AGNs to $\lir$ and only up to
   $\sim 50\%$ contribution by X-ray selected AGN, primarily at
   intermediate specific SFRs. Individual galaxies where AGN
   contribution to $\lir$ is around 90\% are not very common (Figs.\
   \ref{fig:ldustlir_ssfr_agn}--\ref{fig:xrayagn_lir}).

9. \enspace Extreme IR-excess sources similar to those identified at
    $z\sim2$, and possibly related to Compton-thick AGNs, are very
    rare in our sample. Moderate IR excess can be attributed to either
    intermediate-age stellar populations or moderate IR heating from
    AGN (Fig.\ \ref{fig:irx}).

10. \enspace Our findings (items 1--9) are qualitatively the same if
   we compute IR luminosities using \citet{ce01} IR templates instead
   of \citet{dh02}. However, the SED-derived SFRs becomes increasingly
   degenerate with respect to $\lir$ if the $\lir$ is computed using
   \citet{rieke} IR templates (which yield $\lir$ up to an order of
   magnitude higher than \citealt{dh02}). Thus, with \citet{rieke}
   derived $\lir$ item 5 would no longer hold (Fig.\
   \ref{fig:dlir_comp}).

11. \enspace Using a fixed correlation between the UV color (spectral
   slope) and the FUV attenuation to obtain a dust correction, as
   opposed to a correlation that takes into account the effects of SF
   history on UV color, has the effect of producing dust-corrected SFR
   estimates that are on average better correlated with SFR averaged
   over $10^9$ yr, than over $10^8$ yr, i.e., such procedure makes UV
   SFR a poorer indicator of the current SFR (Appendix B).

Our work offers a new approach to study the relation between star
formation and infrared heating. The results would have implications
for a number of studies which use mid-IR luminosity as a tracer of the
current star formation. For example, it could affect the ``time
resolution'' of cosmic SFR densities derived from 24 \mic\ data. Our
results are empirical and are derived from typical data sets used at
intermediate redshifts, but in future work we intend to extend this
study to other redshift regimes and by employing other star formation
indicators.

\acknowledgements

SS would like to thank Christopher N.\ A.\ Willmer, Jeffrey A.\ Newman
and Alison L.\ Coil for help with and access to additional DEEP2 data,
Delphine Marcillac for providing IR templates, and Emanuele Daddi,
David Elbaz, Amelia M.\ Stutz, and Arjun Dey for valuable feedback and
discussions.  This research has made use of NASA's Astrophysics Data
System Bibliographic Services. We acknowledge NSF grants AST-0071198
and AST-0507483 awarded to University of California at Santa Cruz and
Berkeley. This study makes use of data from AEGIS, a multiwavelength
sky survey conducted with the Chandra, GALEX, Hubble, Keck, CFHT, MMT,
Subaru, Palomar, Spitzer, VLA, and other telescopes and supported in
part by the NSF, NASA, and the STFC. This work is based on
observations made with the Spitzer Space Telescope, which is operated
by the Jet Propulsion Laboratory, California Institute of Technology
under a contract with NASA. Support for this work was provided by NASA
through an award issued by JPL/Caltech.


{\it Facilities:} \facility{GALEX}, \facility{Keck:II (DEIMOS)},
\facility{CFHT (MegaPrime/MegaCam)}, \facility{MMT (MegaCam)},
\facility{Hale (WIRC)}, \facility{Spitzer (MIPS)}

\appendix
\section{Robustness of star formation rates from SED fitting}

Many of the results presented in this work depend on UV/optical star
formation rates that we derive using the Bayesian SED fitting. In this
section we evaluate the robustness of our SED fitting technique with
respect to SFRs. We achieve this through simulations in which we try
to recover a {\it known} SFR. In order to make the simulation as
appropriate for our sample as possible, we proceed in the following
manner. Our SED fitting using real data tells us which model in our
library best fits a given real galaxy. So in simulation, we simply
substitute the observed fluxes with corresponding {\it model} fluxes,
but using the {\it observed} flux errors and the redshift. Model
fluxes are scaled to match the observed ones in $i$ band. Then, the
simulated SED fitting proceeds as it would for a real galaxy except
that we exclude from the library the model whose fluxes we are trying
to fit. In this way one gets an exact representation of the SED
fitting for our sample but with the advantage that the {\it true} SFR
(and any other parameter) that one tries to recover is actually known.

\begin{figure*}
\epsscale{1.0} 
\plottwo{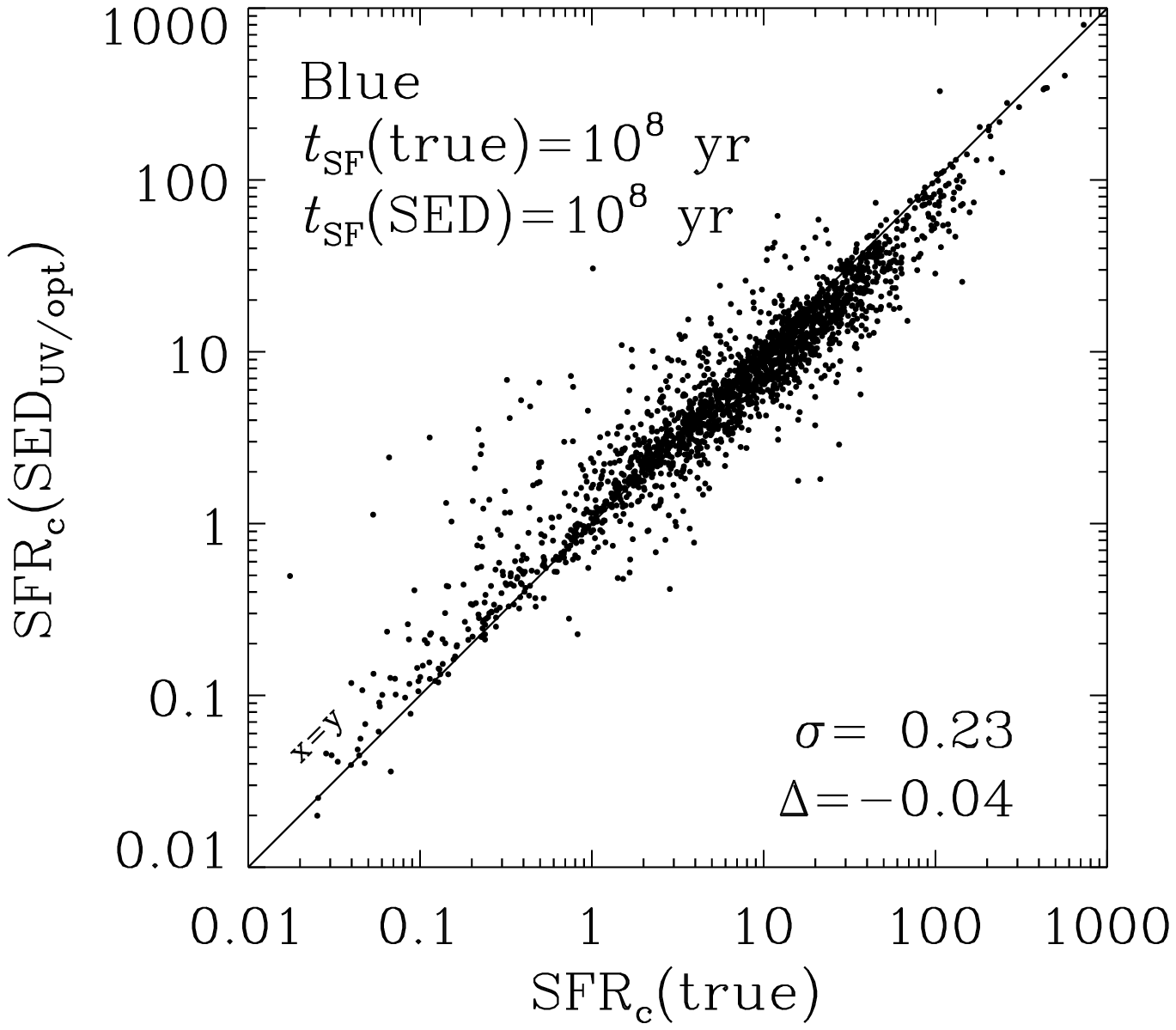}{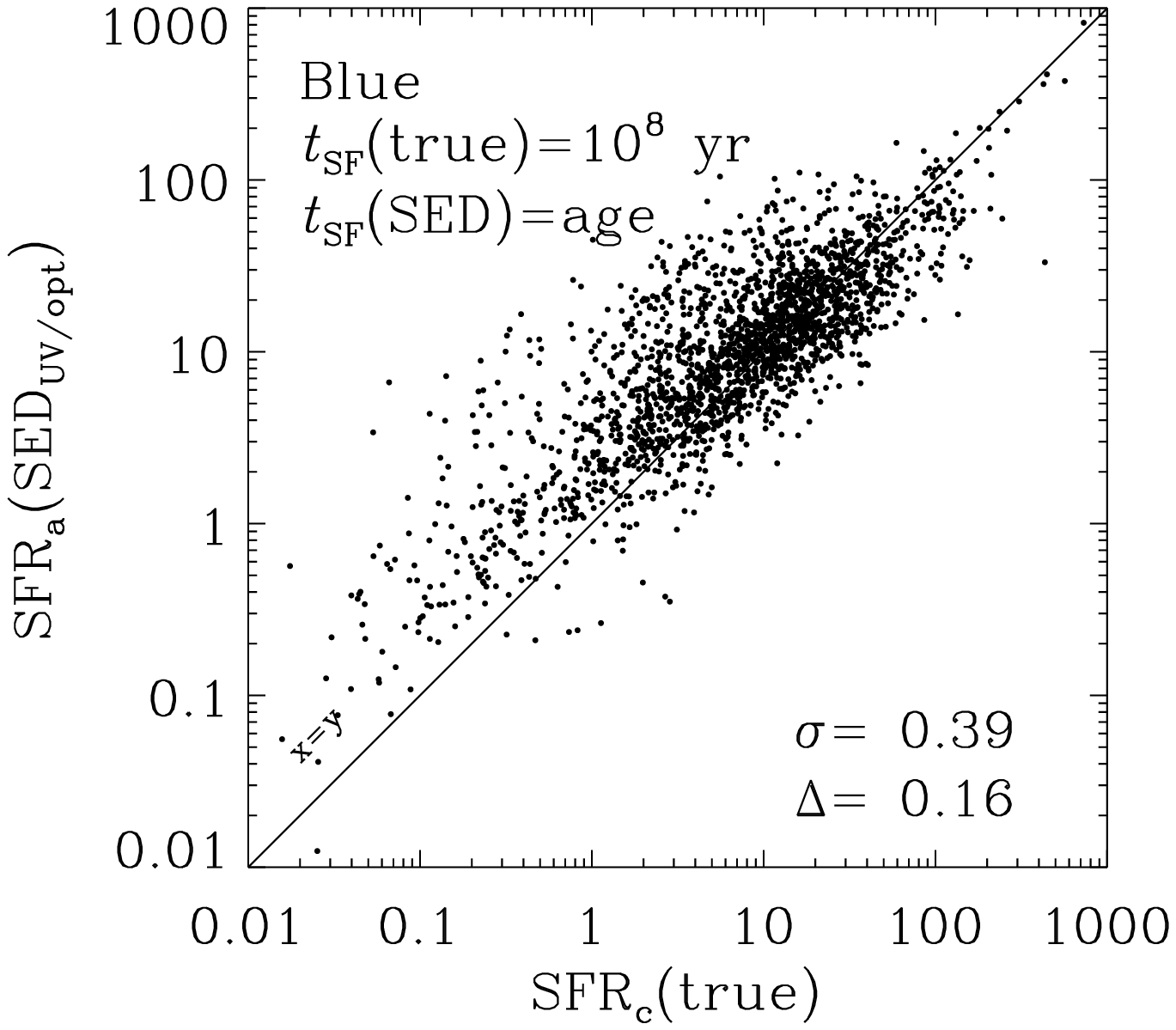}
\caption{Comparison of SFRs recovered from UV/optical SED fitting vs.\
the true current SFRs. The true current SFR is known from the
models. The SED SFR is averaged over $10^8$ yr (left) or over the
population age (1--3 Gyr) (right). As expected, the current SFR is
best recovered by a SFR averaged over a short timescale (left). If
$\lir$ followed the current SFR, one would expect Fig.\
\ref{fig:lir_sfr_blue} and Fig.\ \ref{fig:lir_sfr_t} to respectively
look like the left and the right panel here, flipped around the $x=y$
axis. The fact that they show the opposite behavior argues that $\lir$
does not correlate the best with the current SFR.}
\label{fig:sfr_sim} 
\end{figure*} 

First, we look at how well can the {\it current} SFR be recovered. By
current, in this study we mean the star formation rate averaged over
$t_{\rm SF}=10^8$ yr, the shortest timescale corresponding to
non-ionizing UV emission. In Figure \ref{fig:sfr_sim} {\it left} we
show the comparison of the current SFR retrieved from SED fitting and
the true SFR over the same timescale ($t_{\rm SF}=10^8$ yr). The
objects shown correspond to the sample of blue galaxies detected at 24
\mic\ (i.e., the sample studied in most of \S \ref{sec:lir_sfr}). We
recover the SFRs reasonably well. On average, the SED SFRs fall 0.04
dex below the true ones. The level of discrepancy depends on the SFR
itself. For SFR between 1 and 10 $M_{\odot} {\rm yr}^{-1}$ there is on
average no discrepancy, while for SFR between 10 and 100 $M_{\odot}
{\rm yr}^{-1}$ SFRs from SED fitting are 0.13 dex lower than true
ones. It should not be surprising that systematics at this level are
present. The cause is most likely the limitations in the ability to
obtain the full dynamic range of dust attenuations as discussed in
S07. In any case, the main results in this work rely not on absolute
rates of SF, but on their scatter vs.\ $\lir$. The scatter between SED
SFRs averaged over $10^8$ yr and the true current SFRs is only 0.23
dex. This is very close to what is expected from our estimates of the
{\it individual} errors of SED SFRs (Fig.\ \ref{fig:sfrerr}).

Next we investigate if the derived SED SFR averaged over a timescale
{\it longer} than $10^8$ yr can correlate {\it better} with the
current SFR than the SED SFR averaged over $10^8$ yr itself. This
could possibly be the case if our SED SFRs over $10^8$ yr are simply
more noisy than SFRs averaged over longer timescales. One can imagine
that this could result from the stochastic nature of SF histories in
our models, where bursts have a timescale of roughly $10^8$ yr. In
Figure \ref{fig:sfr_sim} {\it right} we show how SED SFR averaged over
the population age (1--3 Gyr) compares to the true current SFR. The
scatter, 0.39 dex, is considerably worse than in the case where we
averaged SED SFR over the short timescale. Indeed, by checking SED
SFRs averaged over other timescales as well (1 and 2 Gyr), we find
that the current rate of SF is indeed best recovered with the
$10^8$-yr timescale. Additionally, we also find (but do not show in a
plot) that the true SFR averaged over the population age itself is
very well recovered in SED fitting, with a negligible systematic
offset and a scatter of only 0.20 dex.

\section{Dust correction of UV/optical flux using the UV slope} 
\label{sec:uvslope}

To construct the dust-corrected SFRs and UV/optical luminosities we
apply the attenuation model of \citet{cf00} directly to stellar
population models and then compare the reddened models with the
observed SEDs to derive SFRs and other parameters. Thus we use the
full UV/optical SED to constrain the dust attenuation. This procedure
is equivalent to an implicit use of the correlation between the UV
slope (i.e., UV color) and the FUV attenuation \citep{calzetti94}, but
is not identical to it since the \citet{cf00} model (and implicitly
our SED fitting) accounts for the effects of the galaxy SF history on
the UV color \citep{buat}. However, in many instances it is more
practical to perform the dust correction explicitly without
considering the effects of SF history. In those cases one is using
some {\it fixed} correlation between the UV slope and the FUV
attenuation. For our sample of blue-sequence ($^0({\rm NUV}-R)<3.5$)
galaxies we find that the best fixed-slope relation can be fit with

\begin{equation}
A_{FUV} = 3.68\, ^0({\rm FUV}-{\rm NUV})+0.29,\qquad ^0({\rm FUV}-{\rm
NUV})<1, \label{eqn:uvslope}
\end{equation}

\noindent where $^0(FUV-NUV)$ is the rest-frame UV color. This
relation is somewhat steeper than the equivalent relation for local
($z\sim0.1$) SDSS galaxies \citep{s07}, but still not as steep as the
\citet{meurer} relation for local starbursts. While we use the exact
same stellar population and dust models as in S07 and the fit is
constructed in the same way (linear fit through running medians), the
current color cut is somewhat bluer, and more importantly, the two
samples are different, with local galaxies being more quiescent on
average.

Here we would like to draw attention to a systematic effect that, to
our knowledge, has not been discussed elsewhere. Namely, while one
normally expects the unattenuated FUV flux to best correlate with the
SF on timescales of $10^8$ yr, using the fixed-slope relation to
correct the UV flux (i.e., using the same relation between the UV
spectral slope (or color) and the attenuation for {\it all} star
forming galaxies, irrespective of their SF history) will effectively
produce a measure of SF that is {\it on average} somewhat better
correlated with the SF averaged over $10^9$ yr than, as expected, over
$10^8$ yr. For our sample of blue galaxies, we find the scatter around
the linear least-square fit between SFRs (from SED fitting) averaged
over $10^9$ yr and the {\it fixed-slope} dust-corrected FUV luminosity
to be 0.15 dex, compared to 0.17 dex for SFR (from SED fitting)
averaged over $10^8$ yr. On the other hand, as expected, the
correlation of FUV luminosity corrected with {\it full} SED fitting
dust-corrected FUV is the best (0.11 dex) for SFRs averaged over
$10^8$ yr, and significantly worse (0.19 dex) for $10^9$ yr. The
likely explanation for this counter-intuitive effect is that using a
fixed slope between the attenuation and the UV color has the effect of
overestimating the attenuation for galaxies that in reality lie below
that fixed slope, and underestimating it for those that lie above
it. Since galaxies below the fixed slope are more likely to be
galaxies with declining SF, while those above it tend to be more
bursty (e.g., \citealt{kong}), the SFRs of the former get boosted (and
thus become closer to an average over a longer period), while SFRs of
the latter are suppressed, again mimicking the average that includes
the pre-burst period. Because using the fixed slope effectively
lengthens the SF timescale for galaxies with rising or falling SF
histories, it has a consequence (given what we have shown in \S
\ref{sec:lir}) that it will correlate better with the IR luminosity
than the FUV luminosity that was corrected using the full SED
modeling. In other words, using a fixed slope to correct FUV dilutes
our ability to constrain SF over various timescales and study its
relation to the IR.

\begin{deluxetable}{llrrr}
\tablecaption{Summary of data sets} \tablehead{ Survey & Wavelength
range & Total No. & No.\ of galaxies & Survey \\ & or band & of
objects & used in this study\tablenotemark{a} & limit} \startdata
DEEP2 DEIMOS spectroscopy & 6400--9100 \AA & 16087 & 5878 & 24.1
($R_{\rm AB}$) \\ \galex\ Deep Imaging Survey & FUV & 14361 & 1689 &
26.5 (AB) \\ `` & NUV & 54194 & 4363 & 26.5 (AB) \\ MMT $u$ & $u'$ &
71274 & 4807 & 26.3--27.0 (AB)\\ CFHT Legacy Survey & $u^*$ & 367435 &
5438 & 27.2 (AB) \\ `` & $g'$ & 413384 & 5458 & 27.5 (AB)\\ `` & $r'$
& 421258 & 5458 & 27.2 (AB)\\ `` & $i'$ & 426470 & 5458 & 27.0 (AB)\\
`` & $z'$ & 397173 & 5458 & 26.0 (AB)\\ Palomar $K$ & $K_{\rm s}$ &
45008 & 4293 & 21.7--22.5 (AB) \\ {\it Spitzer} MIPS & 24\mic\ & 38049
& 2570 & 30$\mu$Jy \\ \enddata \tablenotetext{a}{Galaxies matched to
DEEP2 spectroscopic data set with (a) secure spectra and (b) lying in
the intersection of CFHTLS and \galex\ coverage (dark gray area in
Figure \ref{fig:area}). Note that a galaxy is kept in the sample
regardless of the presence of a detection in a given UV or
optical/near-IR band.}
\label{tab:data}
\end{deluxetable}

\end{document}